%% file: sap.tex
\documentclass[epjH]{svjour}
\usepackage{graphicx}
\usepackage{amsmath}

\DeclareMathOperator{\tr}{tr}
\newcommand{\nn}{\nonumber}
\newcommand{\be}{\begin{equation}}
\newcommand{\ee}{\end{equation}}
\newcommand{\bea}{\begin{eqnarray}}
\newcommand{\eea}{\end{eqnarray}}

\newcommand{\dotprod} {\raisebox{-.7mm}{\hspace{.4mm}\LARGE$\cdot$\hspace{.4mm}}}  
\newcommand{\crossprod} {\mbox{\boldmath{$\times$}} }

\newcommand{\vecnab}{\mbox{\boldmath{$\nabla$}}}

\newcommand{\bomega}{\mbox{\boldmath{$\omega$}}}

\newcommand{\deer}{(d{\bf r})\,}
\newcounter{saveeqn}%

\begin{document}
\title{Schwinger's Quantum Action Principle. I.}
\subtitle{From Dirac's formulation through Feynman's path integrals to
the Schwinger-Keldysh method}
\author{Kimball A. Milton\inst{1,2}\fnmsep\thanks{\email{milton@nhn.ou.edu}} 
}
\institute{Homer L. Dodge Department of Physics and Astronomy\\
University of Oklahoma\\Norman, OK 73019 USA\thanks{Address 2013--14: 
Laboratoire Kastler Brossel, Universit\'e Pierre
et Marie Curie, 4, place Jussieu Case 74, F-75252 Paris Cedex 05, France}}

\titlerunning{Schwinger's Action Principle}
\authorrunning{K. A. Milton}

\abstract{Starting from the earlier notions of stationary action principles,
we show how Schwinger's Quantum Action Principle descended from Dirac's 
formulation, which independently led Feynman to his path-integral formulation
of quantum mechanics.  The connection between the two is brought out, and
applications are discussed.  The Keldysh-Schwinger time-cycle method of 
extracting matrix elements is described. Part II will discuss the 
variational formulation of quantum electrodynamics and the development of
source. theory.}

\maketitle
\section{Historical Introduction}
\label{intro}
Variational principles for dynamical systems have a long history.  Although
precursors go back at least to Leibnitz (see for example \cite{wikisource})
and Euler \cite{euler}
the ``principle of least action'' was given modern form by de Maupertuis 
\cite{maupertuis1744},
\cite{maupertuis1746}.  We will not attempt to trace the history
here; a brief useful account is given in Sommerfeld's lectures
 \cite{sommerfeld}.  The most important names in the history of the
development of dynamical systems, or at least those that will bear
most directly on the following discussion,   are those of Joseph-Louis Lagrange
\cite{lagrange} and William Rowan Hamilton \cite{Hamilton1834,Hamilton1835}.

Here we are concentrating on the work of Julian Schwinger (1918--1994), who
had profound and pervasive influence on 20th century physics, and whose
many students have become leaders in diverse fields.  For biographical
information about his life and work see \cite{mehra,milton}. Therefore,
we will take up the story in the modern era.
Shortly after Dirac's work with Fock and Podolsky \cite{dfp}, in which the
demonstration of the equivalence between his theory of quantum electrodynamics,
 and that of Heisenberg and Pauli, P. A. M. Dirac wrote a paper on
``The Lagrangian in Quantum Mechanics'' \cite{dirac}. This paper had a profound
influence on Richard Feynman's doctoral dissertation at Princeton on 
``The Principles of Least Action in Quantum Mechanics'' 
\cite{feynman}, and on his
later work on the formulations of the 
``Space-Time Approach to Quantum Electrodynamics'' 
\cite{feynmanst}.  Dirac's paper 
further formed the basis for Schwinger's development of the quantum
action principle, which first appeared in his final operator field
formulation of quantum field theory \cite{Schwinger1951}, 
which we will describe in Part II of this series.

The response of Feynman and Schwinger to Dirac's inspiring paper was completely
different.  Feynman was to give a global ``solution'' to the problem of
determining the transformation function, the probability amplitude connecting
the state of the system at one time to that at a later time, in terms of a sum
over classical trajectories, the famous path integral.  Schwinger, instead,
derived (initially postulated) a differential equation for that transformation
function in terms of a quantum action functional.  This differential equation
possessed Feynman's path integral as a formal solution, which remained
poorly defined; but Schwinger believed throughout his life that his approach
was ``more general, more elegant, more useful, and more tied to the historical
line of development as the quantum transcription of Hamilton's action
principle'' \cite{schwinger1973}.

Later, in a tribute to Feynman, Schwinger commented further.  Dirac, of course,
was the father of transformation theory \cite{Dirac1927}.  
The transformation function from a 
description at time $t_2$ to a description at time $t_1$ is ``the product
 of all the transformations functions associated with the successive
infinitesimal increments in time.''  Dirac said the latter, that is, the
transformation function from time $t$ to time $t+dt$ \textit{corresponds\/}
to $\exp[(i/ \hbar)dt\,L]$, where $L$ is the Lagrangian expressed in terms
of the coordinates at the two times.  For the transformation function between
$t_2$ and $t_1$ ``the integrand is $\exp[(i/\hbar)W]$. where 
$W=\int_{t_2}^{t_1}d t\,L$.''  
``Now we know, and Dirac surely knew, that to within
a constant factor the `correspondence,' for infinitesimal $d t$, 
is an equality
when we deal with a system of nonrelativistic particles possessing a
coordinate-dependent potential energy $V$ \dots.  Why then, did Dirac not make
a more precise, if less general  statement?  Because he was interested in a
general question: What, in quantum mechanics, corresponds to the classical
principle of stationary action?''

``Why, in the decade that followed, didn't someone pick up the computational
possibilities offered by this integral approach to the time transformation
function?  To answer this question bluntly, perhaps no one needed it---until
Feynman came along.'' \cite{Schwinger1989}.

But Schwinger followed the differential route, and starting in early 1950
began a new, his third, formulation of quantum electrodynamics, based on a
variational approach.  This was first published in 1951 \cite{Schwinger1951}. 
A bit later he started developing a new formulation of quantum kinematics,
which he called Measurement Algebra, which got its first public presentation
at \'Ecole de Physique at les Houches in the summer of 1955.
There were several short notes in the Proceedings of the US
National Academy published in 1960, explaining both the quantum kinematical
approach and the dynamical action principle \cite{pnas1}, \cite{pnas2}, 
\cite{pnas3}, \cite{pnas4}, 
but although  he often promised to write a book on the subject 
(as he also promised a book on quantum field theory)
nothing came of it.  Les Houches lectures, based on notes taken by Robert
Kohler, eventually appeared in 1970 \cite{leshouches}. 
Lectures based on a UCLA course by Schwinger were eventually published
under Englert's editorship \cite{Schwinger2001}. 
 The incompleteness
of the written record may be partly alleviated by the present essay. 

We start on a classical footing.

\section{Review of Classical Action Principles}
\label{sec:1}
\input{chap8.tex}
\section{Classical field theory---electrodynamics}
\label{sec:2}
\input{chap9.tex}

\section{Quantum Action Principle}
\label{sec:3}
\input{chap2.tex}


\section{Time-cycle or Schwinger-Keldysh formulation}
\label{sec:5}

A further utility of the action principle is the time-cycle or 
Schwinger-Keldysh formalism, which allows one to calculate matrix elements
and consider nonequilibrium systems.  Schwinger's original work on this
was his famous paper \cite{Schwinger1961}; Keldysh's paper appeared three
years later \cite{Keldysh1964}, and, rather mysteriously, 
cites the Martin-Schwinger equilibrium
paper \cite{Martin1959}, but not the nonequilibrium one \cite{Schwinger1961}.
The following was extracted from notes from Schwinger's lectures given in
1968 at Harvard, as taken by the author.

Consider the expectation value of some physical property $F(t)$ at a particular
time $t_1$ in a state $|b,t_2\rangle$:
\be
\langle F(t_1)\rangle_{b't_2}=\sum_{a'a''}\langle b't_2|a't_1\rangle
\langle a'|F|a''\rangle \langle a''t_1|b't_2\rangle,
\ee
which expresses the expectation value in terms of the matrix elements of the
operator $F$ in a complete set of states defined at time $t_1$, 
$\{|a't_1\rangle\}$.  Suppose the operator $F$ has no explicit time dependence.
Then we can use the action principle to write
\begin{subequations}
\be
\delta \langle a't_1|b't_2\rangle=i\langle a't_1|\delta\left[\int_{t_2}^{t_1}
dt\,L\right]|b't_2\rangle, 
\ee
and so
\be
\delta \langle b't_2|a't_1\rangle=-i\langle b't_2|\delta\left[\int_{t_2}^{t_1}
dt\,L\right]|a't_1\rangle, 
\ee
\end{subequations}
which can be obtained from the first equation by merely exchanging labels,
\be
\int_{t_2}^{t_1}=-\int_{t_1}^{t_2}.
\ee
If we consider
\be
\langle b't_2|b't_2\rangle=\sum_{a'}\langle b't_2|a't_1\rangle\langle a't_1|
b't_2\rangle,
\ee
the above variational equations indeed asserts that
\be
\delta \langle b't_2|b't_2\rangle=0.
\ee

\begin{figure}
\includegraphics{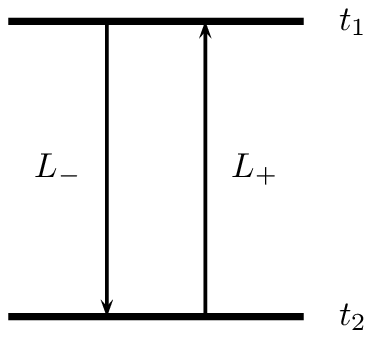}
\caption{\label{fig:tc1} A ``time-cycle,'' in which a system advances forward
in time from time $t_2$ to time $t_1$ under the influence of a Lagrangian
$L_+$, and then backward in time from time $t_1$ back to time $t_2$
under the influence of Lagrangian $L_-$.}
\end{figure}

We can interpret the above as a cycle in time, going from time $t_2$ to $t_1$
and then back again, as shown in Fig,~\ref{fig:tc1}.
  But, now imagine that the dynamics is different on the
forward and return trips, described by different Lagrangians $L_+$ and $L_-$.
Then
\be
\delta\langle b' t_2|b't_2\rangle=i\langle b't_2|\delta\left[\int_{t_2}^{t_1}
dt\,L_+-\int_{t_2}^{t_1}dt \,L_-\right]|b't_2\rangle.  
\ee
In particular, consider a perturbation of the form,
\be H=H_0+\lambda(t)F,\ee
where $\lambda(t)$ is some time-varying parameter.   If
we have an infinitesimal change, and, for example, $\delta
\lambda_+\ne0$, $\delta\lambda_-=0$, then
\be
\delta_{\lambda_+}\langle b't_2|b't_2\rangle^{\lambda_+\lambda_-}
=-i\langle b't_2|\int_{t_2}^{t_1} dt\,\delta \lambda_+F|b' t_2\rangle.
\ee
If we choose  $\delta\lambda_+$ to be an impulse,
\be
\delta\lambda_+=\delta \lambda\delta(t-t'), 
\ee
in this way we obtain the expectation value of $F(t')$.

\begin{figure}
\includegraphics{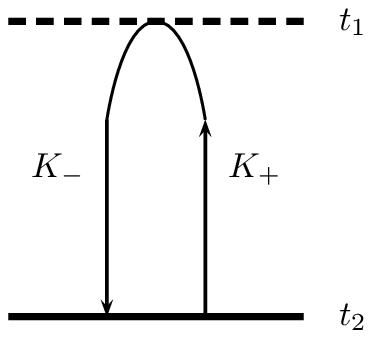}
\caption{\label{fig:tc2} A time cycle in which the harmonic oscillator
evolves from time $t_2$ to time $t_1$ under the influence of a force
$K_+$, and then from $t_1$ back to time $t_2$ under a force $K_-$.}
\end{figure}

Let's illustrate this with a driven harmonic oscillator, as described
by Eq.~(\ref{forcedho}), so now
\begin{subequations}
\bea
H_+&=&\omega y^\dagger y+K_+^*(t)y+K_+(t)y^\dagger,\\
H_-&=&\omega y^\dagger y+K_-^*(t)y+K_-(t)y^\dagger,
\eea
\end{subequations}
which describes the oscillator evolving forward in time from $t_2$ to $t_1$ 
under the influence of the force $K_+$, and backward in time from $t_1$
to $t_2$ under the influence of $K_-$, as shown in Fig.~\ref{fig:tc2}.
From the variational principle we can learn all about $y$ and 
$y^\dagger$.  We have already solved this problem by a more laborious method
above, in Section \ref{sec:chap2}.

It suffices to solve this problem with initial and final ground states,
if we consider only a $K^*$ variation,
\be
\delta_{K^*}\langle 0t_2|0t_2\rangle^{K_+,K_-}=-i\langle 0t|\int_{t_2}^{t_1}
dt\left[\delta K^*_+(t)y_+(t)-\delta K_-^*(t)y_-(t)\right]|0t_2\rangle.
\label{veho}
\ee
Now we must solve the equations of motion, so since effectively $y(t_2)\to0$, 
we have from Eq.~(\ref{first}), 
\begin{subequations}
\bea
y_+(t)&=&-i\int_{t_2}^{t} dt'\,e^{-i\omega(t-t')}K_+(t'),\\
y_-(t)&=&-i\int_{t_2}^{t_1} dt'\,e^{-i\omega(t-t')}K_+(t')
-i\int_{t_1}^{t}dt'\,e^{-i\omega(t-t')}K_-(t').\label{yminust}
\eea
\end{subequations}
The last term in the second equation is
\be
i \int_{t_2}^{t_1}dt'\,e^{-i\omega(t-t')}K(t')\eta(t'-t),
\ee
so naming the advanced and retarded Green's functions by extending
the definition in Eq.~(\ref{hogf}),
\be
G_{a,r}(t,t')=i e^{-i\omega(t-t')}\left\{\begin{array}{c}
\eta(t'-t)\\ -\eta(t-t')\end{array}\right\},
\ee
which satisfy the same differential equation (\ref{diffeq:gf}),
we effectively have
\begin{subequations}
\bea
y_+(t)&=&\int_{t_2}^{t_1}dt'\,G_r(t-t')K_+(t'),\\
y_-(t)&=&-i\int_{t_2}^{t_1} dt'\,e^{-i\omega(t-t')}K_+(t)
+\int_{t_2}^{t_1}dt'\,G_a(t-t')K_-(t'),
\eea
\end{subequations}
The solution to the variational equation (\ref{veho}) is now
\bea
\langle 0t_2|0t_2\rangle^{K_+,K_-}&=&
e^{-i\int dt\,dt'K_+^*(t)G_r(t-t')K_+(t')}
\nn\\
&&\times e^{i\int dt\,dt'K_-^*(t)G_a(t-t')K_-(t')}
e^{\int dt\,dt'K_-^*(t)e^{-i\omega(t-t')}K_+(t')}.\label{tcf}
\eea
This should reduce to 1 when $K_+=K_-=K$, so
\be
-iG_r(t-t')+iG_a(t-t')+e^{-i\omega(t-t')}=0,\label{identity}
\ee
which is, indeed, true.

\begin{figure}
\includegraphics{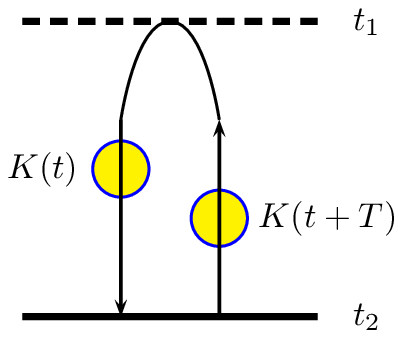}
\caption{\label{fig:tc3} Time cycle in which $K_-(t)=K(t)$, 
$K_+(t)=K(t+T)$, that is, the
forces are the same on the two legs, but displaced in time.}
\end{figure}

As an example, consider $K_-(t)=K(t)$, $K_+(t)=K(t+T)$, that is, the second
source is displaced forward by a time $T$.  
This is sketched in Fig.~\ref{fig:tc3}.
What does this mean?  From a
causal analysis, in terms of energy eigenstates, reading from right to left,
\be
\langle 0t_2|0t_2\rangle^{K_-,K_=}=\sum_n\langle 0t_2|nt_1\rangle^{K_-=K(t)}
\langle nt_1|0t_2\rangle^{K_+=K(t+T)}.
\ee
The effect is the same as moving the $n,t_1$ state to a later time,
\be
\langle nt_1|0t_2\rangle^{K(t+T)}=\langle n t_1+T|0t_2\rangle^{K(t)}
=e^{-in\omega T}\langle nt_1|0t_2\rangle^{K(t)},
\ee
so this says that
\be
\langle 0t_2|0t_2\rangle^{K_-K_+}=\sum_n e^{-in\omega T}p(n,0)^K,
\ee
which gives us the probabilities directly.  From the formula (\ref{tcf}) 
we have, using Eq.~(\ref{identity}), 
\bea
\langle 0t_2|0t_2\rangle^{K_-K_+}&=&
e^{\int dt\,dt'\,K^*(t)e^{-i\omega(t-t')}[K(t'+T)-K(t')]}\nn\\
&=&e^{\int dt\,dt'\,K^*(t)e^{-i\omega(t-t')}[e^{-i\omega T}-1]K(t')]}\nn\\
&=&e^{|\gamma|^2\left(e^{-i\omega T}-1\right)},\label{gfsimple}
\eea
where 
\be
\gamma=\int dt\,e^{i\omega t}K(t).
\ee
Thus we immediately obtain Eq.~(\ref{pnok}), or
\be
p(n,0)^K=e^{-|\gamma|^2}\frac{(|\gamma|^2)^n}{n!}.
\ee

The above Eq.~(\ref{gfsimple})
can be directly used to find certain average values.  For example,
\be
\langle e^{-in\omega T}\rangle^K_0=e^{|\gamma|^2\left(e^{-i\omega T}-1\right)}.
\ee
Expand this for small $\omega T$ and we find
\be
\langle n\rangle_0^K=|\gamma|^2.\label{meanofn}
\ee
In a bit more systematic way we obtain the dispersion:
\be
\langle e^{-i(n-\langle n\rangle)\omega T}\rangle=e^{|\gamma|^2(e^{-i\omega T}
-1+i\omega T)}.\label{expinwt}
\ee
Expanding this to second order in $\omega T$ we get
\be
\langle(n-\langle n\rangle)^2\rangle=\langle n^2\rangle-\langle n\rangle^2
\equiv (\Delta n)^2=|\gamma|^2=\langle n\rangle,\label{nminusn2}
\ee
or
\be
\frac{\Delta n}{\langle n\rangle}=\frac1{\sqrt{\langle n\rangle}}.
\ee
For large quantum numbers, which corresponds to the classical limit, the 
fluctuations become relatively small.

\begin{figure}
\includegraphics{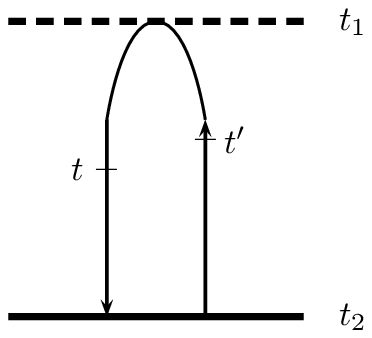}
\caption{\label{fig:tc4} Variational derivatives pick out operators
at definite times $t$ and $t'$.}
\end{figure}

Now consider a more general variational statement than in Eq.~(\ref{veho}),
\be
\delta\langle\,\,|\,\,\rangle^{K_-K_+}=
-i\langle\,\,|\int dt[\delta K_+^*(t)y_+(t)+\dots
-\delta K_-(t)y_-^\dagger-\dots|\,\,\rangle^{K_\pm},
\ee
where the $\dots$ signify the omission of the other source variations, we see
that since we can change the source functions at will, and make very localized
changes,  it makes sense to define the variational derivatives
\begin{subequations}
\bea
&&i\frac\delta{\delta K_+^*(t)}\langle\,\,|\,\,\rangle^{K_\pm}=\langle\,\,|y_+(t)
|\,\,\rangle^{K_\pm},\\
&&-i\frac\delta{\delta K_-(t)}\langle\,\,|\,\,\rangle^{K_\pm}=\langle\,\,|
y_-^\dagger(t)|\,\,\rangle^{K_\pm}.
\eea
\end{subequations}
All expectation values of operator products at any time can be obtained
in this way---in particular, correlation functions.  Repeating this operation
we get
\be
(-i)\frac\delta{\delta K_-(t)}i\frac\delta{\delta K_+^*(t')}\langle t_2|t_2
\rangle^{K_\pm}
=-i\frac\delta{\delta K_-(t)}\langle t_2|y_+(t')|t_2
\rangle^{K_\pm}=\langle t_2|y_-^\dagger(t)y_+(t')|t_2\rangle^{K_\pm}.
\ee
The operators are multiplied in the order of the time development.  The only 
place where $K_-$ appears is in the latter part of the time development.
See Fig.~\ref{fig:tc4}.

The distinction between $\pm$ disappears if we now set $K_+=K_-$:
\be
\frac\delta{\delta K_-(t)}\frac\delta{\delta K_+^*(t')}\langle0 t_2|0t_2
\rangle^{K_\pm}\bigg|_{K_+=K_-=K}=\langle 0t_2|y^\dagger(t)y(t')|0 t_2
\rangle^K.
\ee
As an example, set $t=t'=t_1$; then this reads for the number
operator $N(t)=y^\dagger(t)y(t)$,
\bea
\langle N(t_1)\rangle_0^K&=&\int dt \,K^*(t)G_a(t-t_1) \int dt'G_r(t_1-t')K(t')
\nn\\
&=&i \int dt\, e^{-i\omega(t-t_1)}K^*(t)(-i)\int dt'e^{-i\omega(t_1-t')}K(t')
=|\gamma|^2,
\eea
as before, Eq.~(\ref{meanofn}).

\begin{figure}
\includegraphics{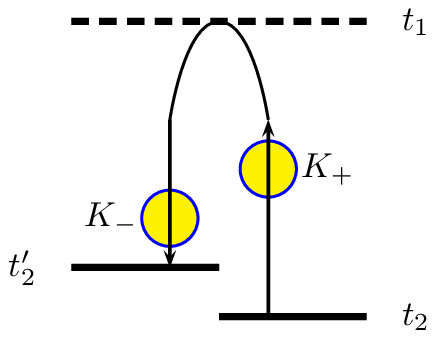}
\caption{\label{fig:tc5} Time cycle with different forces, $K_+$ and $K_-$.
on the forward and backward moving segments.  Now the initial time of the
time cycle, $t_2$, is different from the final time of the time cycle, $t_2'$,
with $\tau=t_2'-t_2$.  It is assumed that the time $t_1$ is later than
both $t_2$ and $t_2'$, and that the forces are localized as shown.}
\end{figure}

We would like to use more general starting and ending states than the ground
state.  We can obtain these by use of impulsive forces.  
It is convenient to deal
with all states at once, as in the generating function for $p(n,0)^K$ 
considered above.  Think of a time cycle starting at time $t_2$, advancing
forward to time $t_1$, during which time the force $K_+$ acts, then moving
back in time to a time $t'_2$, under the influence of the force $K_-$---See
Fig.~\ref{fig:tc5}.
  Let $t_2'=t_2+\tau$.  This displacement injects energy information.
Consider
\be
\sum_n\langle n t_2'|n t_2\rangle^{K_\pm}\equiv \tr \langle t_2'|t_2
\rangle^{K_\pm}=\sum_n e^{-in\omega\tau}\langle nt_2|nt_2\rangle^{K_\pm},
\ee
which uses (no force acts between times $t_2'$ and $t_2$)
\be
\langle n t_2'|=\langle n t_2|e^{-in\omega\tau}.
\ee
Analysis of this formula will yield individual transformation functions.

Now we must solve the dynamical equations subject to boundary conditions.
Let us compare $\tr \langle t_2'|y_+(t_2)|t_2\rangle$ with
$\tr \langle t_2'|y_-(t'_2)|t_2\rangle$.
\begin{subequations}
The first is
\be
\tr \langle t_2'|y_+(t_2)|t_2\rangle=\sum_n\langle n t_2'|y_+(t_2)|n t_2\rangle
=\sum_{nn'}\langle nt_2'|n't_2\rangle\langle n'|y|n\rangle,
\ee
while the second appears as
\be
\tr \langle t_2'|y_-(t_2')|t_2\rangle=\sum_{n'}\langle n' t_2'|y_-(t_2')|n' t_2
\rangle
=\sum_{nn'}\langle n'|y|n\rangle\langle nt_2'|n't_2\rangle.
\ee
\end{subequations}
Here, by introducing a complete set of states at the time of the operator, 
we have expressed the formula in terms 
of the matrix elements of stationary operators.
Remarkably, we see that the two expressions are equal; in effect, there
is a periodicity present here:
\be
y_+(t_2)=y_-(t_2'),\label{periodicity}
\ee
as far as traces are concerned.
Now, the equations of motion (\ref{eom:fho})  for the operators read
\be
\left(i\frac{d}{dt}-\omega\right)y(t)=K(t),
\ee
which has solution (\ref{yminust}) with the addition of the initial term, or
\be
y_-(t)=e^{-i\omega(t-t_2)}y_+(t_2)-i\int_{t_2}^{t_1}dt'\,e^{-i\omega(t-t')}
K_+(t')+i\int_t^{t_1}dt'\,e^{-i\omega(t-t')}K_-(t').
\ee
In particular,
\be
y_-(t_2')=e^{-i\omega\tau}y_+(t_2)-i\int dt'\,e^{-i\omega(t_2+\tau-t')}
(K_+-K_-)(t').\label{inpart}
\ee
Note that the integrals sweep over the full force history.  
Let us let $t_2=0$ for simplicity, although we will keep the label.
Because of the periodicity condition (\ref{periodicity}) this reads
\be
\left(e^{i\omega \tau}-1\right)y_+(t_2)=-i\int dt \, e^{i\omega t}(K_+-K_-)(t)
=-i(\gamma_+-\gamma_-),
\ee
or
\be
y_+(t_2)=\frac1{e^{i\omega \tau}-1}(-i)(\gamma_+-\gamma_-).
\ee

What we are interested in is
\be
\frac{\tr \langle t_2'|t_2\rangle^{K_2}}{\tr \langle t_2'|t_2\rangle},
\ee
The denominator, which refers to the free harmonic oscillator, is immediately
evaluated as
\be
\tr \langle t_2'|t_2\rangle=\sum_{n=0}^\infty e^{-in\omega\tau}
=\frac1{1-e^{-i\omega\tau}}.
\ee
(If $\tau$ be imaginary, we have thermodynamic utility.)  We have then
the variational equation
\be
\delta_{K_\pm^*}\left[\frac{\tr \langle t_2'|t_2\rangle^{K_2}}
{\tr \langle t_2'|t_2\rangle}\right]=
\frac{-i\tr \langle t_2'|\int dt\left(\delta K_+^*y_+-\delta K_-^*y_-\right)
|t_2\rangle^{K_\pm}}{\tr\langle t_2'|t_2\rangle},
\ee

Exactly as before, we get an equation for the logarithm---looking at the 
previous calculation leading to Eq.~(\ref{tcf}), 
we see an additional term, referring to the $y_+(t_2)$
boundary term in Eq.~(\ref{inpart}).  
The periodic boundary condition then gives 
\be
-\frac1{e^{i\omega\tau}-1}\delta(\gamma_+^*-\gamma_-^*)(\gamma_+-\gamma_-).
\ee
Therefore, to convert $\langle 0t_2|0 t_2\rangle^{K_\pm}$ in Eq.~(\ref{tcf}) to
\be
\frac{\tr \langle t_2'|t_2\rangle^{K_2}}{\tr \langle t_2'|t_2\rangle}=
\frac{\sum e^{-in\omega \tau}\langle n t_2|n t_2\rangle^{K_\pm}}{
\sum e^{-in\omega \tau}}
\ee
we must multiply by
\be
\exp[-\frac1{e^{i\omega\tau}-1}|\gamma_+-\gamma_-|^2].
\ee
This holds identically in $\tau$; in particular, in the limit where $\tau
\to -i\infty$, which corresponds to absolute zero temperature, we recover
$\langle 0t_2|0t_2\rangle^{K_\pm}$.

We find, generalizing Eq.~(\ref{tcf})
\bea
&&\frac{\sum_n e^{-in\omega\tau}\langle nt_2|n t_2\rangle^{K_\pm}}
{\sum_n e^{-in\omega\tau}}=e^{-i\int dt\,dt'\,K_+^*(t)G_r(t-t')K_+(t')}\nn\\
&&\times e^{i\int dt\,dt' K_-^*(t) G_0(t-t')K_-(t')}e^{\int dt\,dt'K_-^*(t)
e^{-i\omega(t-t')}K_+(t')}\nn\\
&&\times e^{-(i\omega \tau-1)^{-1}\int dt\,dt'(K_+^*-K_-^*)(t)e^{-i\omega
(t-t')}(K_+-K_-)(t')},\label{complexgf}
\eea
which is the exponential of a bilinear structure.  This is a generating
function for the amplitudes $\langle nt_2|nt_2\rangle^{K_\pm}$.  But it is
useful as it stands.

Put $\tau=-i\beta$; then this describes a thermodynamic average over a thermal
mixture at temperature $T$, where $\beta=1/kT$ in terms of Boltzmann's 
constant.:
\be
\frac{\sum_n e^{-\beta n\omega}\langle\,\,|\,\,\rangle_n}{\sum_n e^{-\beta
n\omega}}
\ee
In terms of this replacement,
\be
\frac1{e^{i\omega\tau}-1}\to \frac1{e^{\beta\omega}-1}=\langle n\rangle_\beta,
\ee
because
\be
\frac{\sum_n n e^{-in\omega\tau}}{\sum_n e^{-in\omega\tau}}=\frac\partial
{\partial(-i\omega\tau)}\ln(\sum_n e^{-in\omega\tau})=
\frac\partial{\partial(-i\omega\tau)}\ln\frac1{1-e^{-i\omega\tau}}
=\frac1{e^{i\omega\tau}-1}.\label{thermal}
\ee

Now consider a time cycle with displacement $T$: the system evolves from time
$t_2$ to time $t_1$ under the influence of the force $K_+(t)$, and backwards
in time from $t_1$ to $t_2'$ under the force $K_-(t)$:
\be
K_-(t)=K(t), \quad K_+(t)=K(t+T).
\ee
This is again as illustrated in Fig.~\ref{fig:tc5}, with these replacements.
What is the physical meaning of this?  Insert in Eq.~(\ref{complexgf})
 a complete set of states at time
$t_1$:
\be
\langle n t_2|n t_2\rangle^{K_\pm}=\sum_{n'}\langle n t_2|n't_1\rangle^{K_-}
\langle n't_1|nt_2\rangle^{K_+}.
\ee
We did this before for the ground state.  The effect is the same as
moving the starting and ending times.
Appearing here is
\be
\langle n't_1|n t_2\rangle^{K(t+T)}=\langle n't_1+T|nt_2+T\rangle^{K(t)}
=e^{-in'\omega T}\langle n't_1|nt_2\rangle^{K(t)}e^{in\omega T}.
\ee
Therefore,
\be
\langle n t_2|nt_2\rangle^{K(t),K(t+T)}=\sum_{n'}e^{-i(n'-n)\omega T}p(n',n)^K
=\langle e^{-i(N-n)\omega T}\rangle_n^K.
\ee
Therefore, as a generalization for finite $\tau$ of 
Eq.~(\ref{expinwt}), we have from Eq.~(\ref{complexgf})
\bea
&&\left(\sum_{n'} e^{-in'\omega \tau}\right)^{-1}\sum_n
 e^{-in\omega\tau}\langle e^{-i(N-n)\omega T}\rangle_n^K\nn\\
&=&\exp\left[\left(e^{-i\omega T}-1\right)|\gamma|^2-\frac1{e^{i\omega\tau}-1}
\left(e^{i\omega T}-1\right)\left(e^{-i\omega T}-1\right)|\gamma|^2\right],
\label{gengf}
\eea
where $T$ gives the final state, and $\tau$ the initial state.  This used
the observation
\be
\int dt\,e^{i\omega t}K(t+T)=e^{-i\omega T}\int dt\,e^{i\omega t}K(t).
\ee
Expand both sides of Eq.~(\ref{gengf}) in powers of $T$, and we learn
\be
-i\omega \sum_n\langle N-n\rangle^K_n \frac{e^{-in\omega \tau}}{\sum_{n'}
e^{-in'\omega\tau}}=-i\omega T|\gamma|^2,\label{angf}
\ee
or
\be
\langle N-n\rangle_\beta^K=|\gamma|^2,\ee
which generalizes an earlier result.  
Now apply Eq.~(\ref{angf}) as a generating function,
\be
\langle N-n\rangle_n^K=|\gamma|^2,
\ee
which reflects the linear nature of the system.

We can rewrite the above generating function more conveniently, by multiplying
by 
\be
e^{i\langle N-n\rangle \omega T}=e^{i\omega T|\gamma|^2},
\ee
that is, Eq,~(\ref{gengf}) can be written as
\bea
&&
\frac1{\sum e^{-in\omega\tau}}\sum e^{-in\omega\tau}\langle e^{-i(N-\langle
N\rangle)\omega T}\rangle_n^K\nn\\
&=&\exp\left[\left(e^{-i\omega T}-1+i\omega T\right)|\gamma|^2-
\frac1{e^{i\omega\tau}-1}\left(e^{-i\omega T}-1\right)\left(e^{i\omega T}-1
\right)|\gamma|^2\right].\label{gf2}
\eea
Now pick off the coefficient of $-(\omega T)^2/2$:
\be
\frac1{\sum e^{-in\omega \tau}}\sum e^{-in\omega \tau}\langle(N-\langle N
\rangle)^2\rangle_n^K=|\gamma|^2+2\frac1{e^{i\omega\tau}-1}|\gamma|^2,
\label{sogf}
\ee
or
\be
\langle (N-\langle N\rangle)^2\rangle_\beta^K=|\gamma|^2[1+2\langle 
n\rangle_\beta].
\ee
If, instead, we multiply Eq.~(\ref{sogf}) 
through by $\sum_n e^{-in\omega\tau}$,
we can use this as a generating function, and learn from Eq.~(\ref{thermal})
that
\be
\langle( N-\langle N\rangle)^2\rangle_n^K=|\gamma|^2(1+2n).
\ee
Note the simplicity of the derivation of this result, which does not involve
complicated functions like Laguerre polynomials.

\section{Prologue}
Let us finally return to the action principle.  Recall from Eq.~(\ref{hovpa})
\be
\langle 0t_1|0t_2\rangle^K=e^{-i\int dt\,dt'K^*(t)G_r(t-t')K(t)}.\label{vpa}
\ee
The action principle says
\be
\delta \langle t_1|t_2\rangle=i\langle t_1|\delta[W_1=\int dt\,L]|t_2\rangle.
\ee
In a general sense, the exponent in Eq.~(\ref{vpa}) is an integrated form
of the action.  In solving the equation of motion, we found in Eq.~(\ref{yoft})
\be
y(t)=e^{-i\omega(t-t_2)}y(t_2)+\int dt' G_r(t-t')K(t'),
\ee
where the first term is effectively zero here.  The net effect is to replace
an operator by a number:
\be
y'(t)=\int dt'G_r(t-t')K(t').
\ee
Then Eq.~(\ref{vpa}) can be written as
\be
\langle 0t_1|0t_2\rangle^K=e^{-i\int dt\,K^*(t)y'(t)}.\label{vpap}
\ee
Recall that the action was was the integral of the
Lagrangian (\ref{holagrange}), or
\be
W=\int dt\left[y^\dagger i\frac\partial{\partial t}y-\omega y^\dagger y
-y^\dagger K(t)-yK^*(t)\right],
\ee
so we see one term in Eq.~(\ref{vpap}) here, and the equation of motion 
(\ref{eom:fho}) cancels
out the rest!  So let's add something which gives the equation for $y'$:
\be
\langle 0t_1|0t_2\rangle^K=e^{i\int dt\left[y^{\dagger\prime}i\frac{d}{dt}y'
-\omega y^{\dagger\prime}y'-y^{\dagger\prime}K-y'K^*\right]}=e^{iW}.
\ee
Now insist that $W$ is stationary with respect to variations of $y'$, 
$y^{\dagger\prime}$, and we recover the equation of motion,
\be
\left(i\frac{d}{dt}-\omega\right)y'(t)=K(t).
\ee
This is the starting point for the development of source theory, which will
be treated in Part II.



\section{End of Part I}
\label{concl}
We have traced Schwinger's development of action formulations from classical
systems of particles and fields, to the description of quantum dynamics 
through the Quantum Action Principle.  In the latter, we here described only
quantum mechanical systems, especially the driven harmonic oscillator.
This is ahistorical, since Schwinger first developed his quantum dynamical
principle in the context of quantum electrodynamics in the early 1950s, and 
only nearly a decade later applied it to quantum mechanics, which is field
theory in one dimension---time.  At roughly the same time he was thinking
about quantum statistical systems \cite{Martin1959}, and it was natural to
turn to a description of nonequilibrium systems, which was the motivation
of the time-cycle method, although Schwinger put it in a general, although
simplified, context.
The time cycle method was immediately applied to quantum field theory by
his students, K. T. Mahanthappa and P. M. Bakshi
 \cite{mahanthappa1962,bakshi1963}. But rather
than here tracing the profound and growing influence of this great paper,
as well as the deep underpinning still provided by  Schwinger's action
principle, 
we need to carry out a sketch of the application of these methods
to quantum field theory, and to what Schwinger perceived as the successor to
field theory, Source Theory.  But we have now reached a appropriate point to
pause.  In Part II of this paper we will provide that elaboration, and trace
some of the vast influence that Schwinger's development of these powerful
techniques have had in all branches of theoretical physics.
  
\begin{acknowledgement}
I thank the Laboratoire Kastler Brossel, ENS, UPMC, CNRS, for its hospitality
during the completion of this manuscript.  I especially thank Astrid 
Lambrecht and Serge Reynaud. The work was completed in part
with funding from the Simons Foundation and the CNRS.  I thank my many students
at the University of Oklahoma, where much of the material reported here
was used as the basis of lectures in quantum mechanics and quantum field 
theory.
\end{acknowledgement}
%
%

\end{document}

%% file: chap8.tex
This section is based on Chapter 8 of {\it Classical Electrodynamics} 
\cite{Schwinger1998}, a substantially transformed version of lectures
given by Schwinger at UCLA around 1974.  (Remarkably, he never gave
lectures on this subject at Harvard after 1947.)

\begin{figure}
\centering
\includegraphics{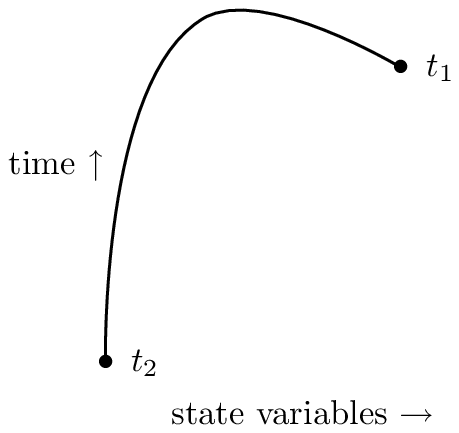}
\caption{\label{fig8.1} A possible path from initial state to final state.}
\end{figure}

      We start by reviewing and generalizing the Lagrange-Hamilton principle
 for a single particle. The action, $W_{12}$, is defined as the time
 integral of the Lagrangian, $L$, where the integration extends from an initial
 configuration or state at time $t_2$ to a final state at time $t_1$:
\begin{equation}
W_{12}=\int_{t_2}^{t_1} dt\, L. 
\label{8.1}
\end{equation}
The integral refers to any path, any line of time development, from the
initial to the final state, as shown in Fig.\ \ref{fig8.1}.
The actual time evolution of the system is selected by the principle
of stationary action: In response to infinitesimal variations of the
integration path, the action $W_{12}$ is stationary---does not have
a corresponding infinitesimal change---for variations about the correct
path, provided the initial and final configurations are held fixed,
\begin{equation}  
\delta W_{12} =0.  
\label{8.3}
\end{equation}
This means that, if we allow infinitesimal changes at the initial and
final times, including alterations of those times, the only contribution
to $\delta W_{12}$ then comes from the endpoint variations, or
 \begin{equation}
 \delta W_{12} = G_1 - G_2,
\label{8.2}
\end{equation}
where $G_a$, $a=1$ or $2$,  is a function, called the generator,
 depending on dynamical variables only at time $t_a$.
 In the following, we will consider three different realizations of
 the action principle, where, for simplicity, we will restrict our 
 attention to a single particle.

\subsection{Lagrangian Viewpoint}

      The nonrelativistic motion of a particle of mass $m$ moving in a 
potential $V({\bf r},t)$ is described by the Lagrangian
\begin{equation}
 L={1\over2}m\left({d{\bf r}\over dt}\right)^2-V({\bf r},t).
\label{8.4}
\end{equation}
 Here, the independent variables are $\bf r$ and $t$, so that two kinds
of variations can be considered.  First, a particular motion is
altered infinitesimally, that is,  the path is changed by
an amount $\delta{\bf r}$:
 \begin{equation}
{\bf r}(t) \to {\bf r}(t) + \delta {\bf r}(t).
\label{8.6}
\end{equation}
 Second, the final and initial times can be altered infinitesimally, by
$\delta t_1$ and $\delta t_2$, respectively.
It is more convenient, however, to think of these time displacements
as produced by a continuous variation of the time parameter, $\delta t(t)$,
\begin{equation}
 t \to t + \delta t(t),
\label{8.5}
\end{equation}
 so chosen that, at the endpoints,
\begin{equation}
 \delta t(t_1) = \delta t_1,\qquad    \delta t(t_2) = \delta t_2.
\label{1-18.5}
\end{equation}
 The corresponding change in the time differential is
\begin{equation}
 dt \to d (t + \delta t) = \left(   1 + {d\delta t\over dt }\right)dt,
\label{8.7}
\end{equation}
 which implies the transformation of the time derivative,
\begin{equation}
{ d\over dt}\to \left(1-{ d\delta t \over   dt}\right){d\over dt}.
\label{8.8}
\end{equation}
 Because of this redefinition of the time variable, the limits of integration
 in the action,
 \begin{equation}
 W_{12}=\int_2^1\left[{1\over2}m{(d{\bf r})^2\over dt}-dt\,V\right],
 \label{1-18.8}
 \end{equation}
   are {\it not\/} changed, the time displacement being produced through
   $\delta t(t)$ subject to (\ref{1-18.5}).
 The resulting variation in the action is now
\begin{eqnarray}
\delta W_{12}&=& \int_2^1dt\left\{m{ d{\bf r}\over dt}\dotprod {d \over dt}
\delta{\bf r}-\delta{\bf r}\dotprod\vecnab V   -{d\delta t\over dt}
\left[{1\over2}m\left({d{\bf r}\over dt}\right)^2+V\right]
-\delta t{\partial\over\partial t}V\right\}
\nonumber\\
&=&\int_2^1 dt\Bigg\{{d\over dt}\left[m{d{\bf r}\over dt}\dotprod \delta{\bf r}
-\left({1\over 2}m\left({d{\bf r}\over dt}\right)^2+V\right)\delta t\right]
\nonumber\\
&&\!\!\!\!\!\mbox{}+\delta{\bf r}\dotprod\left[-m{d^2\over dt^2}{\bf r}-\vecnab V
\right]+\delta t\left({d\over dt}
\left[{1\over2}m\left({d{\bf r}\over dt}\right)^2+V
\right]-{\partial\over\partial t}V\right)\!\!\!\Bigg\},
\label{8.9}
\end{eqnarray}
 where, in the last form, we have shifted the time derivatives in order to
 isolate $\delta{\bf r}$ and $\delta t$.
 
    Because $\delta{\bf r}$ and $\delta t$ are independent variations, 
the  principle of stationary action implies that the actual motion 
is governed by
\begin{subequations}
\begin{eqnarray}
m{ d^2\over dt^2}{\bf r}=&-&\vecnab V,
\label{8.10}\\
{d\over dt}\bigg[{1\over2} m \left({ d{\bf r}\over dt}\right)^2    &+& V
\bigg]={\partial\over\partial t}V,
\label{8.11}
\end{eqnarray}
 while the total time derivative gives the change at the endpoints,
\begin{equation}
      G = {\bf p\dotprod \delta r} - E\delta t,
\label{8.12}
\end{equation}
 with
\begin{equation}
\mbox{momentum}={\bf p}=m{d{\bf r}\over dt},\qquad
\mbox{energy}= E={1\over 2}m\left({ d{\bf r}\over dt}\right)^2+V.
\end{equation}
\end{subequations}
 Therefore, we have derived Newton's second law [the equation of motion in
 second-order form], (\ref{8.10}), 
and, for a static potential, $\partial V/\partial t=0$,
 the conservation of energy, (\ref{8.11}). The
 significance of (\ref{8.12}) will be discussed later in Section 
\ref{sec:1}.\ref{8.4}.

 \subsection{Hamiltonian Viewpoint}
      Using the above definition of the momentum, we can rewrite the Lagrangian
as
\begin{equation}
 L={\bf p}\dotprod{d{\bf r}\over   dt}- H({\bf r}, {\bf p},t),
\label{8.13}
\end{equation}
 where we have introduced the Hamiltonian
\begin{equation}
 H={ p^2\over 2m} + V ({\bf r},t).
\label{8.14}
\end{equation}
 We are here to regard $\bf r$, $\bf p$, and $t$ as independent variables in
 \begin{equation}
 W_{12}=\int_2^1[{\bf p}\dotprod d{\bf r}-dt\, H].
 \label{1-18.18}
 \end{equation}
 The change in the action, when $\bf r$, $\bf p$, and $t$ are all varied, is
\begin{eqnarray}
 \delta W_{12}
&=&\int_2^1 dt\left[{\bf  p}\dotprod{d\over dt}\delta{\bf r}-
\delta{\bf r}\dotprod{\partial H\over\partial{\bf r}}+\delta{\bf p}
\dotprod{d{\bf r}\over dt}-\delta{\bf p}\dotprod{\partial H\over\partial{\bf p}}
-{d\delta t\over dt}H-\delta t{\partial H\over\partial t}\right]\nonumber\\
&=&\int_2^1dt\bigg[{d\over dt}({\bf p\dotprod\delta r}-H\delta t)
+\delta{\bf r}\dotprod\left(-{d{\bf p}\over dt}-{\partial H\over\partial{\bf r}}
\right)\nonumber\\
&&\quad+\delta{\bf p}\dotprod\left({d{\bf r}\over dt}-{\partial H\over 
\partial {\bf p}}\right)+\delta t\left({dH\over dt}-{\partial H
\over\partial t}\right)\bigg].
\label{8.15}
\end{eqnarray}
 The action principle then implies
\begin{subequations}
\begin{eqnarray}
{d{\bf r}\over dt}&=&{\partial H\over\partial{\bf p}}={{\bf p}\over m},
\label{8.16}\\
{d{\bf p}\over dt}&=&-{\partial H\over \partial {\bf r}}=-\vecnab V,
\label{8.17}\\
{d H\over dt}&=&{\partial H\over\partial t},\label{8.18}\\
G&=&{\bf p\dotprod\delta r}-H\delta t.
\label{8.19}
\end{eqnarray}
\end{subequations}
In contrast with the Lagrangian differential equations of motion, which
involve second derivatives, these Hamiltonian equations contain only
first derivatives; they are called first-order equations.
They describe the same physical system, because when 
 (\ref{8.16}) is substituted into (\ref{8.17}), we recover 
 the Lagrangian-Newtonian equation (\ref{8.10}).
 Furthermore, if we insert (\ref{8.16}) into the Hamiltonian
 (\ref{8.14}), we identify $H$ with $E$.  The third equation
 (\ref{8.18}) is then identical with (\ref{8.11}).  We also note
 the equivalence of the two versions of $G$.
 
 But probably the most direct way of seeing that the same physical
 system is involved comes by writing the Lagrangian in the Hamiltonian
 viewpoint as
 \begin{equation}
 L={m\over2}\left({d{\bf r}\over dt}\right)^2-V-{1\over2m}
 \left({\bf p}-m{d{\bf r}\over dt}\right)^2.
 \label{1-18.26}
 \end{equation}
 The result of varying $\bf p$ in the stationary action principle
 is to produce
 \begin{equation}
 {\bf p}=m{d{\bf r}\over dt}.
 \label{1-18.27}
 \end{equation}
 But, if we accept this as the {\em definition\/} of $\bf p$, the
 corresponding term in $L$ disappears and we explicitly regain the 
 Lagrangian description.  We are justified in completely omitting the
 last term on the right side of (\ref{1-18.26}), despite its
 dependence on the variables $\bf r$ and $t$, because of its quadratic
 structure.  Its explicit contribution to $\delta L$ is
 \begin{equation}
 -{1\over m}\left({\bf p}-m{d{\bf r}\over dt}\right)\dotprod
 \left(\delta{\bf p}-m{d\over dt}\delta {\bf r}+m{d{\bf r}\over dt}
 {d\delta t\over dt}\right),
 \label{1-18.28}
 \end{equation}
 and the equation supplied by the stationary action principle for
 $\bf p$ variations, (\ref{1-18.27}), also guarantees that there is no
 contribution here to the results of $\bf r$ and $t$ variations.
 
 \subsection{A Third, Schwingerian, Viewpoint}
\label{sec8.3}

      Here we take $\bf r$, $\bf p$, and the velocity,
$\bf v$, as independent variables, so that
 the Lagrangian is written in the form
\begin{equation}
    L =  {\bf p}\dotprod\left({d{\bf r}\over dt}-{\bf v}\right)
           + {1\over2}  mv^2 - V({\bf r},t)\equiv
{\bf p}\dotprod{d{\bf r}\over dt}-H({\bf r,p,v},t),
\label{8.20}
\end{equation}
where 
\begin{equation}
H({\bf r,p,v},t)={\bf p\dotprod v}-{1\over2}mv^2+V({\bf r},t).
\label{8.26}
\end{equation}
 The variation of the action is now
\begin{eqnarray}
\delta W_{12}&=&\delta\int_2^1[{\bf p}\dotprod d{\bf r}-H\,dt]\nonumber\\
&=&\int_2^1dt\bigg[\delta{\bf p}\dotprod{d{\bf r}\over dt}
+{\bf p}\dotprod{d\over dt}\delta{\bf r}
-\delta{\bf r}\dotprod{\partial H\over\partial{\bf r}}
-\delta{\bf p}\dotprod{\partial H\over\partial{\bf p}}
-\delta{\bf v}\dotprod{\partial H\over\partial {\bf v}}\nonumber\\
&&\qquad\mbox{}-\delta t{\partial H\over\partial t}-H{d\delta t\over dt}\bigg]
\nonumber\\
&=&\int_2^1dt\bigg[{d\over dt}({\bf p\dotprod\delta r}- H\delta t)
-\delta{\bf r}\dotprod\left({d{\bf p}\over dt}+{\partial H\over\partial{\bf r}}
\right)\nonumber\\
&&\mbox{}+\delta{\bf p}\dotprod\left({d{\bf r}\over dt}-{\partial H
\over\partial {\bf p}}\right)
-\delta{\bf v}\dotprod{\partial H\over \partial{\bf v}}
+\delta t\left({d H\over dt}
-{\partial H\over\partial t}\right)\bigg],
\label{8.21}
\end{eqnarray}
 so that the action principle implies
\begin{subequations}
\begin{eqnarray}
{d{\bf p}\over dt}&=&-{\partial H\over\partial{\bf r}}=-\vecnab V,
\label{8.22}\\
{d{\bf r}\over dt}&=&{\partial H\over\partial{\bf p}}={\bf v},
\label{8.23}\\
{\bf 0}&=&-{\partial H\over\partial {\bf v}}=-{\bf p}+m{\bf v},
\label{8.24}\\
{d H\over dt}&=&{\partial H\over\partial t},
\label{8.25}\\
G&=&{\bf p\dotprod\delta r}-H\delta t.
\label{8.27}
\end{eqnarray}
\end{subequations}
 Notice that there is no equation of motion for $\bf v$ since $d{\bf v}/dt$
 does not occur in the Lagrangian, nor is it multiplied by a time derivative.
 Consequently, (\ref{8.24}) refers to
 a single time and is an equation of constraint.

      From this third approach, we have the option of returning to either of
the other two viewpoints by imposing an appropriate restriction.  Thus, if
we write (\ref{8.26}) as
\begin{equation}
H({\bf r,p,v},t)={p^2\over 2m}+V({\bf r},t)-{1\over2m}({\bf p}-m{\bf v})^2,
\label{1-18.33}
\end{equation}
and we adopt
\begin{equation}
{\bf v}={1\over m}{\bf p}
\label{1-18.34}
\end{equation}
as the {\em definition\/} of $\bf v$, we recover the Hamiltonian
description, (\ref{8.13}) and (\ref{8.14}). 
 Alternatively, we can present the Lagrangian (\ref{8.20}) as
\begin{equation}
L={m\over2}\left(d{\bf r}\over dt\right)^2-V+({\bf p}-m{\bf v})
\dotprod\left({d{\bf r}\over dt}-{\bf v}\right)-{m\over2}
\left({d{\bf r}\over dt}-{\bf v}\right)^2.
\label{1-18.35}
\end{equation}
Then, if we adopt the following as {\em definitions},
\begin{equation}
{\bf v}  = {d{\bf r}\over dt},\quad {\bf p}=m{\bf v},
\label{18.36}
\end{equation}
the resultant form of $L$ is that of the Lagrangian viewpoint, (\ref{8.4}).
It might seem that only the definition ${\bf v}=d{\bf r}/dt$, inserted
in (\ref{1-18.35}), suffices to regain the Lagrangian description.
But then the next to last term in (\ref{1-18.35}) would give the
following additional contribution to $\delta L$, associated with
the variation $\delta {\bf r}$:
\begin{equation}
({\bf p}-m{\bf v})\dotprod{d\over dt}\delta {\bf r}.
\label{1-18.37}
\end{equation}

      In the next Section, where the action formulation of electrodynamics is
 considered, we will see the advantage of adopting this third approach, which
 is characterized by the introduction of additional variables,
 similar to $\bf v$, for which there
 are no equations of motion.

 \subsection{Invariance and Conservation Laws}
 \label{sec8.4}
 There is more content to the principle of stationary action than equations
 of motion.  Suppose one considers a variation such that
 \begin{equation}
 \delta W_{12}=0,
 \label{1-19.1}
 \end{equation}
 independently of the choice of initial and final times.  We say that the
 action, which is left unchanged, is {\em invariant\/} under this alteration
 of path.  Then the stationary action principle (\ref{8.2}) asserts that
 \begin{equation}
 \delta W_{12}=G_1-G_2=0,
 \label{1-19.2}
 \end{equation}
 or, there is a quantity $G(t)$ that has the same value for any choice of
 time $t$; it is conserved in time.  A differential statement of that
 is
 \begin{equation}
 {d\over dt}G(t)=0.
 \label{1-19.3}
 \end{equation}
 The $G$ functions, which are usually
 referred to as generators, express the interrelation between 
conservation laws and invariances of the system.  
 
 Invariance implies conservation, and vice versa. 
 A more precise statement is the following:
\begin{quote}
      If there is a conservation law, the action is stationary under an
      infinitesimal transformation in an appropriate variable.
\end{quote}
 The converse of this statement is also true.
\begin{quote}
 If the action $W$ is invariant under an infinitesimal transformation (that is,
$\delta W = 0$), then there is a corresponding conservation law.
\end{quote}
  This is the celebrated theorem proved by Amalie Emmy Noether 
\cite{Noether1918}.
  
  Here are some examples.
 Suppose the Hamiltonian of (\ref{8.13}) does not depend explicitly
 on time, or
 \begin{equation}
 W_{12}=\int_2^1[{\bf p}\dotprod d{\bf r}-H({\bf r,p})dt].
 \label{1-19.4}
 \end{equation}
 Then the variation (which as a rigid displacement in time, amounts
 to a shift in the time origin)
 \begin{equation}
 \delta t=\mbox{constant}
 \label{1-19.5}
 \end{equation}
 will give $\delta W_{12}=0$ [see the first line of (\ref{8.15}),
 with $\delta{\bf r}=0$, $\delta{\bf p}=0$, $d\delta t/dt=0$, 
$ \partial H/\partial t=0$].  The conclusion is that $G$ in (\ref{8.19}),
which here is just
\begin{equation}
G_t=-H\delta t,
\label{1-19.6}
\end{equation}
is a conserved quantity, or that
\begin{equation}
{dH\over dt}=0.
\label{1-19.7}
\end{equation}
This inference, that the Hamiltonian---the energy---is conserved, if there is
no explicit time dependence in $H$, is already present in (\ref{8.18}).
But now a more general principle is at work.

\begin{figure}
\centering
\includegraphics{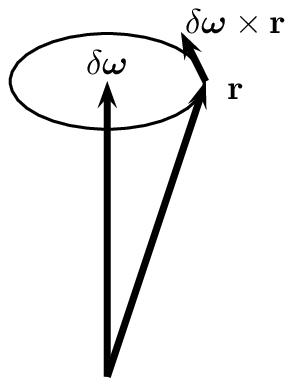}
\caption{$\delta\mbox{\boldmath{$\omega$}}\crossprod{\bf r}$ is perpendicular
to $\delta\mbox{\boldmath{$\omega$}}$ and $\bf r$, and represents an
infinitesimal rotation
of $\bf r$ about the $\delta\mbox{\boldmath{$\omega$}}$ axis.}
\label{fig8.2}
\end{figure}

Next, consider an infinitesimal, rigid rotation, one that maintains the
lengths and scalar products of all vectors.  Written explicitly for the
position vector $\bf r$, it is
\begin{equation}
 \delta{\bf r}  = \delta\mbox{\boldmath{$\omega$}}\crossprod{\bf r},
\end{equation}
where the constant vector $\delta\mbox{\boldmath{$\omega$}}$ gives 
the direction 
and magnitude of the rotation (see Fig.\ \ref{fig8.2}).
Now specialize (\ref{8.14}) to
\begin{equation}
H={p^2\over2m}+V(r),
\label{1-19.9}
\end{equation}
where $r=|{\bf r}|$, a rotationally invariant structure.  Then
\begin{equation}
W_{12}=\int_2^1[{\bf p}\dotprod d{\bf r}-H\, dt]
\label{1-19.10}
\end{equation}
is also invariant under the rigid rotation, implying the conservation of
\begin{equation}
G_{\delta\mbox{\boldmath{$\omega$}}}={\bf p\dotprod\delta r=
\delta\mbox{\boldmath{$\omega$}}\dotprod r\crossprod p.}
\label{1-19.11}
\end{equation}
This is the conservation of angular momentum,
\begin{equation}
{\bf L = r\crossprod p}, \quad {d\over dt}{\bf L=0}.
\label{1-19.12}
\end{equation}
 Of course, this is also contained within the equation of motion,
\begin{equation}
{ d \over dt}{\bf L}= -{\bf r} \crossprod\vecnab V=
-{\bf r\crossprod\hat r}{\partial V\over\partial r}={\bf 0},
\end{equation}
 since $V$ depends only on $|{\bf r}|$.

Conservation of linear momentum appears analogously when there is
invariance under a rigid translation.  For a single particle, (\ref{8.17})
tells us immediately that $\bf p$ is conserved if $V$ is a constant, say
zero.  Then, indeed, the action
\begin{equation}
W_{12}=\int_2^1\left[{\bf p}\dotprod d{\bf r}-{p^2\over2m}dt\right]
\label{1-19.13}
\end{equation}
is invariant under the displacement
\begin{equation}
\delta{\bf r}=\delta\mbox{\boldmath{$\epsilon$}}=\mbox{constant},
\label{1-19.14}
\end{equation}
and
\begin{equation}
G_{\delta\mbox{\boldmath{$\epsilon$}}}={\bf p}\dotprod
\delta\mbox{\boldmath{$\epsilon$}}
\label{1-19.15}
\end{equation}
is conserved.  But the general principle acts just as easily for, say,
a system of two particles, $a$ and $b$, with Hamiltonian
\begin{equation}
H={p_a^2\over2m_a}+{p_b^2\over2m_b}+V({\bf r}_a-{\bf r}_b).
\label{1-19.16}
\end{equation}
This Hamiltonian and the associated action
\begin{equation}
W_{12}=\int_2^1[{\bf p}_a\dotprod d{\bf r}_a+{\bf p}_b\dotprod d{\bf r}_b
-H\,dt]
\label{1-19.17}
\end{equation}
are invariant under the rigid translation
\begin{equation}
\delta {\bf r}_a=\delta{\bf r}_b=\delta\mbox{\boldmath{$\epsilon$}},
\label{1-19.18}
\end{equation}
with the implication that
\begin{equation}
G_{\delta\mbox{\boldmath{$\epsilon$}}}={\bf p}_a\dotprod\delta{\bf r}_a
+{\bf p}_b\dotprod\delta{\bf r}_b=({\bf p}_a+{\bf p}_b)\dotprod
\delta\mbox{\boldmath{$\epsilon$}}
\label{1-19.19}
\end{equation}
is conserved.  This is the conservation of the total linear momentum,
\begin{equation}
{\bf P}={\bf p}_a+{\bf p}_b,\quad {d\over dt}{\bf P=0}.
\label{1-19.20}
\end{equation}

Something a bit more general appears when we consider a rigid translation
that grows linearly in time:
\begin{equation}
\delta{\bf r}_a=\delta{\bf r}_b=\delta{\bf v}\,t,
\label{1-19.21}
\end{equation}
using the example of two particles.  This gives each particle the common
additional velocity $\delta{\bf v}$, and therefore must also change
their momenta,
\begin{equation}
\delta{\bf p}_a=m_a\delta {\bf v},\quad\delta{\bf p}_b=m_b\delta {\bf v}.
\label{1-19.22}
\end{equation}
The response of the action (\ref{1-19.17}) to this variation is
\begin{eqnarray}
\delta W_{12}&=&\int_2^1[({\bf p}_a+{\bf p}_b)\dotprod\delta{\bf v}\,dt
+\delta{\bf v}\dotprod(m_a d{\bf r}_a+m_b d{\bf r}_b)-({\bf p}_a
+{\bf p}_b)\dotprod\delta{\bf v}\,dt]\nonumber\\
&=&\int_2^1d[(m_a{\bf r}_a+m_b{\bf r}_b)\dotprod\delta{\bf v}].
\label{1-19.23}
\end{eqnarray}
The action is {\em not\/} invariant; its variation has end-point contributions.
But there is still a conservation law, not of $G={\bf P\dotprod\delta v}t$,
but of ${\bf N}\dotprod\delta{\bf v}$, where
\begin{equation}
{\bf N=P}t-(m_a{\bf r}_a+m_b{\bf r}_b).
\label{1-19.24}
\end{equation}
Written in terms of the center-of-mass position vector
\begin{equation}
{\bf R}={m_a{\bf r}_a+m_b{\bf r}_b\over M},\quad M=m_a+m_b,
\label{1-19.25}
\end{equation}
the statement of conservation of
\begin{equation}
{\bf N=P}t-M{\bf R},
\label{1-19.26}
\end{equation}
namely
\begin{equation}
{\bf 0}={d{\bf N}\over dt}={\bf P}-M{d{\bf R}\over dt},
\label{1-19.27}
\end{equation}
is the familiar fact that the center of mass of an isolated system
moves at the constant velocity given by the ratio of the total momentum
to the total mass of that system.

\subsection{Nonconservation Laws. The Virial Theorem}

The action principle also supplies useful nonconservation laws.  Consider,
for constant $\delta\lambda$,
\begin{equation}
\delta{\bf r}=\delta\lambda{\bf r},\quad \delta {\bf p}=-\delta\lambda
{\bf p},
\label{1-19.28}
\end{equation}
which leaves ${\bf p}\dotprod d{\bf r}$ invariant,
\begin{equation}
\delta({\bf p}\dotprod d{\bf r})=(-\delta\lambda{\bf p})\dotprod d{\bf r}
+{\bf p}\dotprod(\delta\lambda d{\bf r})=0.
\label{1-19.29}
\end{equation}
But the response of the Hamiltonian
\begin{equation}
H=T(p)+V({\bf r}),\quad T(p)={p^2\over2m},
\label{1-19.30}
\end{equation}
is given by the noninvariant form
\begin{equation}
\delta H=\delta\lambda(-2T+{\bf r}\dotprod\vecnab V).
\label{1-19.31}
\end{equation}
Therefore we have, for an arbitrary time interval, for the
variation of the action (\ref{1-18.18}),
\begin{equation}
\delta W_{12}=\int_2^1 dt[ \delta\lambda(2T-{\bf r}\dotprod\vecnab V)]
=G_1-G_2=\int_2^1 dt{d\over dt}({\bf p\dotprod\delta\lambda r})
\label{1-19.32}
\end{equation}
or, the theorem
\begin{equation}
{d\over dt}{\bf r\dotprod p}=2T-{\bf r}\dotprod\vecnab V.
\label{1-19.33}
\end{equation}

For the particular situation of the Coulomb potential between charges,
$V=\mbox{constant}/r$, where
\begin{equation}
{\bf r}\dotprod\vecnab V=r{d\over dr}V=-V,
\label{1-19.34}
\end{equation}
the virial theorem asserts that
\begin{equation}
{d\over dt}({\bf r\dotprod p})=2T+V.
\label{1-19.35}
\end{equation}
We apply this to a {\em bound\/} system produced by a force of attraction.
On taking the time average of (\ref{1-19.35}) the time derivative term
disappears.  That is because, over an arbitrarily long time interval
$\tau=t_1-t_2$, the value of ${\bf r\dotprod p}(t_1)$ can differ by only
a finite amount from ${\bf r\dotprod p}(t_2)$, and
\begin{equation}
\overline{{d\over dt}({\bf r\dotprod p})}\equiv{1\over\tau}\int_{t_2}^{t_1}
dt{d\over dt}{\bf r\dotprod p}=
{{\bf r\dotprod p}(t_1)-{\bf r\dotprod  p}(t_2)\over\tau}\to0,
\label{1-19.36}
\end{equation}
as $\tau\to\infty$. The conclusion, for time averages,
\begin{equation}
2\overline T=-\overline V,
\label{1-19-37}
\end{equation}
is familiar in elementary discussions of motion in a $1/r$ potential.

Here is one more example of a nonconservation law:  Consider the
variations
\begin{subequations}
\begin{eqnarray}
\delta{\bf r}&=&\delta\lambda{{\bf r}\over r},\\
\delta{\bf p}&=&-\delta\lambda\left({{\bf p}\over r}-{{\bf r\, p\dotprod
r}\over r^3}\right)=\delta\lambda{{\bf r\crossprod(r\crossprod p)}
\over r^3}.
\label{1-19.38}
\end{eqnarray}
\end{subequations}
Again ${\bf p}\dotprod d{\bf r}$ is invariant:
\begin{equation}
\delta({\bf p}\dotprod d{\bf r})=-\delta\lambda\left({{\bf p}\over r}-
{{\bf r \,p\dotprod r}\over r^3}\right)\dotprod d{\bf r}
+{\bf p}\dotprod\left(\delta\lambda{d{\bf r}\over r}-\delta\lambda
{\bf r}{{\bf r}\dotprod d{\bf r}\over r^3}\right)=0,
\label{1-19.39}
\end{equation}
and the change of the Hamiltonian (\ref{1-19.30}) is now
\begin{equation}
\delta H=\delta\lambda\left[-{{\bf L}^2\over mr^3}+{{\bf r}\over r}\dotprod
\vecnab V\right].
\label{1-19.40}
\end{equation}
The resulting theorem, for $V=V(r)$, is
\begin{equation}
{d\over dt}\left({{\bf r}\over r}\dotprod {\bf p}\right)=
{{\bf L}^2\over mr^3}-{dV\over dr},
\label{1-19.41}
\end{equation}
which, when applied to the Coulomb potential, gives the bound-state time
average relation
\begin{equation}
{L^2\over m}\overline{\left({1\over r^3}\right)}=-\overline{\left(V\over r
\right)}.
\label{1-19.42}
\end{equation}
This relation is significant in hydrogen fine-structure calculations
(for example, see \cite{Schwinger2001}).







%% file: chap9.tex

This section is based on Chapter 9 of \textit{Classical Electrodynamics},
\cite{Schwinger1998}, which again in turn grew, torturously,  
out of Schwinger's UCLA lectures. Here we use Gaussian units.

\subsection{Action of Particle in Field}

It was stated in our review of mechanical action principles in the
previous section that the
third viewpoint, which employs the variables $\bf r$, $\bf p$, and $\bf v$,
was particularly convenient for describing electromagnetic forces on
charged particles.  With the explicit, and linear, appearance of
$\bf v$ in what plays the role of the potential function 
when magnetic fields are present, 
we begin to see the basis for that remark.  Indeed, we have only to consult
(\ref{8.20}) to find the appropriate Lagrangian:
\begin{equation}
L={\bf p}\dotprod\left({d{\bf r}\over dt}-{\bf v}\right)+{1\over2}mv^2-e\phi
+{e\over c}{\bf v\dotprod A},
\label{1-20.16}
\end{equation}
where $\phi$ and $\mathbf{A}$ are the scalar and vector potentials,
respectively.
To recapitulate, the equations resulting from variations of $\bf p$, $\bf r$,
and $\bf v$ are, respectively,
\begin{subequations}
\begin{eqnarray}
{d{\bf r}\over dt}&=&{\bf v},\label{1-20.17a}\\
{d\over dt}{\bf p}&=&-e\vecnab \left[\phi-{1\over c}{\bf v\dotprod A}\right],
\label{1-17.20b}\\
{\bf p}&=&m{\bf v}+{e\over c}{\bf A}.
\label{1-20.17c}
\end{eqnarray}
\end{subequations}

We can now move to either the Lagrangian or the Hamiltonian formulation.
For the first, we simply adopt ${\bf v}=d{\bf r}/dt$ as a definition
(but see the discussion in Sec.~\ref{sec8.3}) and get
\begin{equation}
L={1\over2}m\left(d{\bf r}\over dt\right)^2-e\phi+{e\over c}{d{\bf r}\over dt}
\dotprod{\bf A}.
\label{1-20.18}
\end{equation}
Alternatively, we use (\ref{1-20.17c}) to define
\begin{equation}
{\bf v}={1\over m}\left({\bf p}-{e\over c}{\bf A}\right),
\label{1-20.19}
\end{equation}
and find
\begin{subequations}
\begin{eqnarray}
L&=&{\bf p}\dotprod{d{\bf r}\over dt}-H,\\
H&=&{1\over2m}\left({\bf p}-{e\over c}{\bf A}\right)^2+e\phi.
\label{1-20.20}
\end{eqnarray}
\end{subequations}

\subsection{Electrodynamic Action}

The electromagnetic field is a mechanical system.  It contributes its variables to the
action, to the Lagrangian of the whole system of charges and fields.  
In contrast with the point charges, the field is distributed in space.  
Its Lagrangian should therefore be, not a summation over discrete points, 
but an integration over all spatial volume elements,
\begin{equation}
L_{\rm field}=\int\deer{\cal L}_{\rm field};
\label{1-21.1}
\end{equation}
this introduces the Lagrange function, or Lagrangian density, ${\cal L}$.
The total Lagrangian must be the sum of  the particle part, (\ref{1-20.16}),
 and the field part, (\ref{1-21.1}), where the latter must be chosen so as 
 to give the Maxwell equations in Gaussian units: 
\begin{subequations}
 \begin{eqnarray}
\vecnab\crossprod{\bf B}&=&{1\over c}{\partial\over\partial t}{\bf E}+
{4\pi\over c}{\bf j},\quad \vecnab\dotprod{\bf E}=4\pi\rho,\\
-\vecnab\crossprod{\bf E}&=&{1\over c}{\partial\over \partial t}{\bf B},
\!\quad\quad\qquad\vecnab\dotprod{\bf B}=0.
\end{eqnarray}
\end{subequations}
 The homogeneous equations here are
equivalent to the construction of the electromagnetic
field in term of potentials, or, 
\begin{subequations}
\begin{eqnarray}
{1\over c}{\partial\over\partial t}{\bf A}&=&-{\bf E}-\vecnab\phi,\\
{\bf B}&=&\vecnab\crossprod{\bf A}.
\label{1-21.2}
\end{eqnarray}
\end{subequations}
Thus, we recognize that  ${\bf A(r},t)$, ${\bf E(r},t)$, in analogy with
${\bf r}(t)$, ${\bf p}(t)$, obey equations of motion while $\phi({\bf r},t)$,
${\bf B(r},t)$, as analogues of ${\bf v}(t)$, do not.  There are enough clues
here to give the structure of ${\cal L}_{{\rm field}}$, apart from an overall
factor.  The anticipated complete Lagrangian for microscopic
 electrodynamics is
\begin{eqnarray}
L &=&\sum_a \left[{\bf p}_a\dotprod\left({d{\bf r}_a\over dt} - {\bf v}_a
\right)   + {1\over2} m_av_a^2-e_a\phi({\bf r}_a)
+{e_a\over c}{\bf v}_a\dotprod{\bf A(r}_a)\right]\nonumber\\
&&\!\!\!\!\!\!\!\!\!\!\mbox{}+{1\over4\pi}\int(d{\bf r})\,
\left[{\bf E}\dotprod
\left(-{1\over c}{\partial\over\partial t}{\bf A}-\vecnab\phi\right)
-{\bf B\dotprod\vecnab \crossprod A} + {1\over2} (B^2- E^2) \right].
\label{9.1}
\end{eqnarray}

 The terms that are summed in (\ref{9.1}) describe the behavior of charged 
 particles under the influence of the fields, while the terms that are integrated 
 describe the field behavior.  The independent variables are
\begin{equation} 
{\bf r}_a(t), \quad {\bf v}_a(t),\quad {\bf p}_a(t),\quad  \phi({\bf r},t),
\quad {\bf A}({\bf r},t),\quad {\bf E}({\bf r},t),\quad {\bf  B}({\bf r},t),
\quad t.
\end{equation}
 We now look at the response of the Lagrangian to variations in each of these
 variables separately, starting with the particle part:
\begin{subequations}
\begin{eqnarray}
 \delta{\bf r}_a:\quad \delta L &=& {d\over dt} (\delta{\bf r}_a\dotprod{\bf p}_a)
   + \delta{\bf r}_a\dotprod\left[ - {d{\bf p}_a\over dt}-\vecnab_ae_a\left(
 \phi({\bf r}_a) -{{\bf v}_a\over c}\dotprod{\bf A(r}_a) \right)\right],\nonumber\\
\label{9.2}\\
 \delta{\bf v}_a:\quad \delta L &=& \delta{\bf v}_a\dotprod\left[
-{\bf p}_a+m_a{\bf v}_a+{e_a\over c}{\bf A(r}_a)\right],
\label{9.3}\\
 \delta{\bf p}_a: \quad\delta L&=& \delta{\bf p}_a \dotprod\left(
{ d{\bf r}_a\over  dt} - {\bf v}_a\right).
\label{9.4}
\end{eqnarray}
\end{subequations}
 The stationary action principle now implies the equations of motion
\begin{subequations}
\begin{eqnarray}
{ d{\bf p}_a\over dt}&=&-e_a\vecnab_a\left(\phi({\bf r}_a)-{{\bf v}_a\over c}
\dotprod {\bf A(r}_a)\right),
\label{9.5a}\\
 m_a{\bf v}_a&=&{\bf p}_a- { e_a\over c}{\bf A (r}_a),
\label{9.5b}\\
{\bf v}_a&=&{d{\bf r}_a\over dt},
\label{9.5c}
\end{eqnarray}
\end{subequations}
 which are the known results, (\ref{1-20.17a})--(\ref{1-20.17c}).

The real work now lies in deriving the equations of motion for the fields.
 In order to cast all the field-dependent terms into integral form, we 
introduce charge and current densities,
\begin{subequations}
\begin{eqnarray}
\rho({\bf r},t)&=&\sum_a e_a\delta({\bf r-r}_a(t)),\\
{\bf j(r},t)&=&\sum_a e_a {\bf v}_a(t)\delta({\bf r-r}_a(t)),
\label{1-22.3}
\end{eqnarray}
\end{subequations}
 so that
\begin{equation}
\sum_a\left[-e_a\phi({\bf r}_a)+{e_a\over c}{\bf v}_a\dotprod{\bf A(r}_a)
\right]=\int(d{\bf r})\,\left[-\rho\phi+{1\over c}{\bf j\dotprod A}\right].
\label{9.6}
\end{equation}
 The volume integrals extend over sufficiently large regions to contain all
 the fields of interest.  Consequently, we can integrate by parts and ignore
 the surface terms.  The responses of the Lagrangian (\ref{9.1}) 
 to field variations, and the corresponding equations of motion deduced 
 from the action principle are
\begin{subequations}
\label{edvareqn}
\begin{eqnarray}
\delta\phi:\qquad \delta L&=&{1\over4\pi}\int(d{\bf r})\,\delta\phi
(\vecnab\dotprod{\bf E}-4\pi\rho),
\label{9.7a}\\
 \vecnab\dotprod{\bf E} &=& 4\pi\rho,
\label{9.7b}\\
 \delta{\bf A}:\qquad \delta L&=&-{1\over4\pi c}{d\over dt}\int
(d{\bf r})\, \delta{\bf A}\dotprod {\bf E} \nonumber\\
&&\mbox{} + {1\over4\pi}
\int\deer\delta{\bf A}\dotprod\left({1\over c}{\partial{\bf E}\over \partial t}
+{4\pi\over c}{\bf j}-\vecnab \crossprod{\bf B}\right),
\label{9.8a}\\
\vecnab\crossprod {\bf B}&=&{1\over c}{\partial\over\partial t}{\bf E}+
{4\pi\over c}{\bf j},
\label{9.8b}\\
 \delta{\bf E}:\qquad \delta L&=&{1\over4\pi}\int\deer\delta{\bf E}\dotprod\left(
-{1\over c}{\partial\over\partial
 t}{\bf A}-\vecnab\phi-{\bf E}\right),
\label{9.9a}\\
{\bf E}&=&-{1\over c}{\partial\over\partial t}{\bf A}-\vecnab\phi,
\label{9.9b}\\
 \delta{\bf B}:\qquad \delta L&=&{1\over4\pi}\int\deer\delta{\bf B}
\dotprod(-\vecnab\crossprod {\bf A+B}),
\label{9.10a}\\
{\bf B}&=&\vecnab   \crossprod {\bf A}.
\label{9.10b}
\end{eqnarray}
\end{subequations}
 We therefore recover Maxwell's equations, two of which are implicit in the
 construction of ${\bf E}$ and $\bf B$ in terms of potentials.  
By making a time variation of the action [variations due to the time dependence
 of the fields vanish by virtue of the stationary action principle---that is,
 they are already subsumed in Eqs.~(\ref{edvareqn}),
\begin{equation}
\delta t:\qquad\delta W=\int dt\,\left[
 {d \over dt}  (-H\delta t) + \delta t {dH\over dt} \right],
\label{9.11}
\end{equation}
 we identify the Hamiltonian of the system to be
\begin{eqnarray}
 H &=& \sum_a\left[\left({\bf  p}_a -{e_a\over c}{\bf A(r}_a)\right)\dotprod 
 {\bf v}_a - {1\over2}m_av_a^2+ e_a\phi({\bf r}_a)\right]\nonumber\\
&&+{1\over4\pi}\int\deer\left[{\bf E}\dotprod\vecnab\phi + {\bf B}\dotprod
\vecnab\crossprod {\bf A}+{1\over2}(E^2-B^2)\right],
\label{9.12}
\end{eqnarray}
 which is a constant of the motion,
$dH/dt=0$.
 The generators are inferred from
 the total time derivative terms in (\ref{9.2}), (\ref{9.8a}), and 
(\ref{9.11}),
\begin{subequations}
\begin{equation}
\delta W_{12}=G_1-G_2,
\end{equation}
 to be
\begin{equation}
 G = \sum_a\delta{\bf r}_a\dotprod
{\bf p}_a-{1\over4\pi c}\int\deer{\bf E \dotprod\delta A }- H\delta t.
\label{9.13}
\end{equation}
\end{subequations}

\subsection{Energy}
Notice that the total Lagrangian (\ref{9.1}) can be presented as
\begin{equation}
L=\sum_a{\bf p}_a\dotprod{d{\bf r}_a\over dt}-{1\over4\pi c}\int\deer
{\bf E}\dotprod{\partial\over\partial t}{\bf A}-H,
\label{1-23.1}
\end{equation}
where the Hamiltonian is given by (\ref{9.12}).
The narrower, Hamiltonian, description is reached by eliminating all
variables that do not obey equations of motion, and, correspondingly,
do not appear in $G$.  Those ``superfluous'' variables are the ${\bf v}_a$
and the fields $\phi$ and $\bf B$, which are eliminated by using
  (\ref{9.5b}), (\ref{9.7b}), and (\ref{9.10b}), 
 the equations without time derivatives, resulting, first,
  in the intermediate form
\begin{equation}
 H = \sum_a\left({1\over2m_a}\left({\bf p}_a-{e_a\over c}{\bf A}({\bf r}_a)
 \right)^2 
+e_a\phi({\bf r}_a)\right)+ \int\deer \left[ {E^2+B^2\over8\pi}-\rho\phi\right].
\label{9.14}
\end{equation}
 The first term here is the energy of the particles moving in the field 
[particle energy---see (\ref{1-20.20})], so we might call the second term the 
field energy.  
The ambiguity of these terms (whether the potential energy of particles 
is attributed to them or to the fields, or to both) is evident from 
the existence of a simpler form of the Hamiltonian
\begin{equation}
 H = \sum_a{1\over2 m_a}\left({\bf p}_a-{e_a\over c}{\bf A}({\bf r}_a)\right)^2
+\int\deer {E^2 + B^2\over8\pi},
\label{9.15}
\end{equation}
 where we have used the equivalence of the two terms involving 
$\phi$, given in (\ref{9.6}). 

This apparently startling result suggests that the scalar potential has 
disappeared from the dynamical description.  But, in fact, it has not.  
If we vary the Lagrangian (\ref{1-23.1}), where $H$ is given by 
(\ref{9.15}), with respect to $\bf E$ we find
\begin{equation}
\delta L=-{1\over4\pi}\int\deer\delta {\bf E}\dotprod\left({1\over c}
{\partial\over\partial t}{\bf A+E}\right)=0.
\label{1-23.9}
\end{equation}
Do we conclude that ${1\over c}{\partial\over\partial t}{\bf A+E=0}$?
That would be true if the $\delta{\bf E(r},t)$ were arbitrary.  They are not;
${\bf E}$  is  subject to the restriction---the constraint---(\ref{9.7b}), which
means that any change in $\bf E$ must obey
\begin{equation}
\vecnab\dotprod\delta{\bf E}=0.
\label{1-23.10}
\end{equation}
The proper conclusion is that the vector multiplying $\delta{\bf E}$ in
(\ref{1-23.9}) is the gradient of a scalar function, just as in (\ref{9.9b}),
\begin{equation}
{1\over c}{\partial\over\partial t}{\bf A+E}=-\vecnab\phi,
\label{1-23.11}
\end{equation}
for that leads to
\begin{equation}
\delta L=-{1\over4\pi}\int\deer(\vecnab\dotprod\delta{\bf E})\phi=0,
\label{1-23.12}
\end{equation}
as required.

The fact that the energy is conserved,
\begin{equation}
      {dH\over dt} = 0,
\label{9.16}
\end{equation}
where
\begin{equation}
H=\sum_a{1\over2}m_av_a^2+\int\deer U,\quad U={E^2+B^2\over8\pi},
\label{1-24.2}
\end{equation}
is a simple sum of particle kinetic energy and integrated field energy
density,  can be verified directly by
 taking the time derivative of (\ref{9.14}). 
The time rate of change of the particle energy is computed directly:
\begin{equation}
{d\over dt}\sum_a\left({1\over2}m_av_a^2+e_a\phi({\bf r}_a)\right)
=\sum_a{\partial\over\partial t}\left(e_a\phi({\bf r}_a)
-{e_a\over c}{\bf v}_a\dotprod{\bf A(r}_a)\right).
\label{9.17}
\end{equation}    
 We can compute the time derivative of the field energy by using the equation
 of energy conservation, 
\begin{equation}
{d\over dt}\int\deer U=-\int\deer {\bf j\dotprod E},
\label{9.18}
\end{equation}
 to be
\begin{eqnarray}
{d\over dt}\int\deer\left({ E^2 +B^2\over8\pi}-\rho\phi\right)&=&
\int\deer\left[-{\bf j \dotprod E}-\phi{\partial\over\partial t}\rho
-\rho{\partial\over\partial t}\phi\right]\nonumber\\
&=&-\int\deer\left[\rho{\partial\over\partial t}\phi-{1\over c}{\bf j}
\dotprod {\partial\over\partial t}{\bf A}\right]\nonumber\\
&=&-\sum_a e_a\left({\partial\over\partial t}\phi({\bf r}_a)-{1\over c}
{\bf v}_a\dotprod{\partial\over\partial t}{\bf A(r}_a)\right).\nonumber\\
\label{9.19}
\end{eqnarray}
 Here we have used (\ref{9.9b}), and have noted that
\begin{equation}
\int\deer\left[{\bf j}\dotprod\vecnab\phi-\phi{\partial\over\partial t}
\rho\right]=0
\end{equation}
 by charge conservation.  Observe that (\ref{9.17}) and 
(\ref{9.19}) are equal in magnitude
 and opposite in sign, so that their sum is zero.  
This proves the statement of energy conservation (\ref{9.16}).

\subsection{Momentum and Angular Momentum Conservation}
      The action principle not only provides us with the field equations,
 particle equations of motion, and expressions for the energy, but also with
 the generators (\ref{9.13}). The generators provide a connection between 
conservation laws and invariances of the action (recall Section \ref{sec8.4}). 
 Here we will further illustrate this connection by deriving momentum and 
angular momentum conservation from the invariance of the action under 
rigid coordinate translations and rotations, respectively. 
[In a similar way we could derive energy conservation, (\ref{9.16}), 
from the invariance under time displacements---see also Section \ref{sec9.7}].

\begin{figure}
\centering
\includegraphics{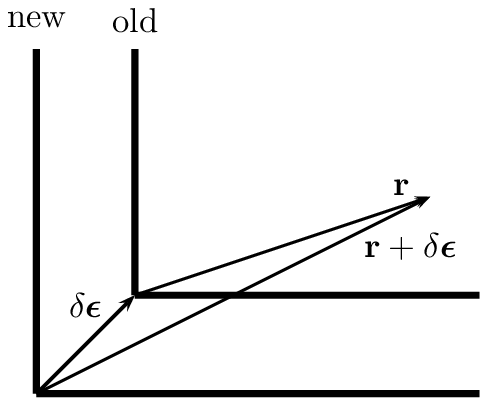}
\caption{\label{fig9.1} Rigid coordinate displacement, where
the new coordinate system is displaced by a rigid translation 
$-\delta\mbox{\boldmath{$\epsilon$}}$ relative to the old coordinate system.}
\end{figure}

 Under an infinitesimal rigid coordinate displacement, 
 $\delta{\mbox{\boldmath{$\epsilon$}}}$, 
a given point which is described by $\bf r$ in the old coordinate 
system is described by ${\bf r}+\delta{\mbox{\boldmath{$\epsilon$}}}$
in the new one. (See Fig.\ \ref{fig9.1}.)
 The response of the particle term in (\ref{9.13}) is simple: 
$\delta {\mbox{\boldmath{$\epsilon$}}} \dotprod\sum_a{\bf p}_a$; for the
field part, we require the change, $\delta {\bf A}$, 
of the vector potential induced by
 the rigid coordinate displacement. The value of a field ${\cal F}$ 
at a physical point $P$ is unchanged under such a displacement,
so that if $\bf r$ and $\bf r+\delta \mbox{\boldmath{$\epsilon$}}$
 are the coordinates of $P$ in the two frames, there are corresponding
functions $F$ and $\overline F$ such that
\begin{equation}
{\cal F}(P) = F({\bf r})  =\overline{F}( {\bf r}  + 
\delta {\mbox{\boldmath{$\epsilon$}}}),
\label{9.20}
\end{equation}
 that is, the new function $\overline{F}$
 of the new coordinate equals the old function
 $F$ of the old coordinate.  The change in the function $F$ at the {\it same 
coordinate\/} is given by
\begin{equation}
\overline{F}({\bf r}) = F({\bf r}) + \delta F({\bf r}),
\end{equation}
 so that
\begin{equation}
 \delta F({\bf r}) = F({\bf r}-\delta {\mbox{\boldmath{$\epsilon$}}}) 
 - F({\bf r}) = 
-\delta{\mbox{\boldmath{$\epsilon$}}}\dotprod\vecnab F({\bf r}),
\label{9.21}
\end{equation}
 for a rigid translation (not a rotation).  
 
 As an example, consider the charge density
 \begin{equation}
 \rho({\bf r})=\sum_a e_a\delta({\bf r-r}_a).
 \label{1-24.7}
 \end{equation}
 If the positions of all the particles, the ${\bf r}_a$, are displaced by
 $\delta{\mbox{\boldmath{$\epsilon$}}}$, the charge density changes to
 \begin{equation}
 \rho({\bf r})+\delta\rho({\bf r})=\sum_a e_a\delta({\bf r-r}_a-
 \delta{\mbox{\boldmath{$\epsilon$}}}),
 \label{1-24.8}
 \end{equation}
 where
\begin{equation}
 \delta({\bf r-r}_a-\delta{\mbox{\boldmath{$\epsilon$}}})=\delta({\bf r-r}_a)
 -\delta{\mbox{\boldmath{$\epsilon$}}}\dotprod\vecnab_r\delta({\bf r-r}_a),
 \label{1-24.9}
 \end{equation}
 and therefore
 \begin{equation}
 \delta\rho({\bf r})=-\delta{\mbox{\boldmath{$\epsilon$}}}
 \dotprod\vecnab\rho({\bf r}),
 \label{1-24.10}
 \end{equation}
 in agreement with (\ref{9.21}).
 
  So the field part of $G$ in (\ref{9.13}) is
\begin{eqnarray}
-\int(d{\bf r})\,{1\over4\pi c} {\bf E \dotprod \delta A}&=&{1\over4\pi c}
\int\deer E_i(\delta{\mbox{\boldmath{$\epsilon$}}}
\dotprod \vecnab)A_i\nonumber\\
=&-&{1\over c}\sum_a e_a\delta {\mbox{\boldmath{$\epsilon$}}}
\dotprod{\bf A(r}_a)+
{1\over4\pi c}\int\deer({\bf E\crossprod B)}\dotprod\delta 
{\mbox{\boldmath{$\epsilon$}}},
\label{9.22}
\end{eqnarray}
 where the last rearrangement makes use of (\ref{9.7b}) and (\ref{9.10b}),
 and the vector identity
 \begin{equation}
 \delta {\mbox{\boldmath{$\epsilon$}}}\crossprod(\vecnab\crossprod {\bf A})
 =\vecnab(\delta {\mbox{\boldmath{$\epsilon$}}}\dotprod{\bf A})
 -(\delta {\mbox{\boldmath{$\epsilon$}}}\dotprod\vecnab){\bf A}.
 \end{equation}
Including the particle part from (\ref{9.13}) we find the generator
corresponding to a rigid coordinate displacement can be written as
\begin{equation}
G=\delta{\bf {\mbox{\boldmath{$\epsilon$}}}\dotprod P},
\label{9.23}
\end{equation}
 where
\begin{equation}
{\bf P}=\sum_a\left({\bf p}_a-{e_a\over c}{\bf A(r}_a)\right)
+{1\over4\pi c}\int\deer{\bf E \crossprod B}\equiv\sum_a m_a{\bf v}_a+
\int\deer {\bf G},
\label{9.24}
\end{equation}
 with $\bf G$ the momentum density. 
Since the action is invariant under a rigid displacement,
\begin{equation}
 0 = \delta W = G_1 - G_2 = ({\bf P}_1 -{\bf P}_2)\dotprod\delta{\bf r},
\end{equation}
 we see that
\begin{equation}
 {\bf P}_1 =  {\bf P}_2,
\end{equation}
 that is, the total momentum, $\bf P$, is conserved.  
This, of course, can also be verified by explicit calculation:
\begin{eqnarray}
{d\over dt}\int\deer {1\over4\pi c}{\bf E \crossprod B}
&=&-\int\deer \left[\rho{\bf E}+{1\over c}{\bf j\crossprod B}\right]\nonumber\\
&=&-\sum_ae_a\left({\bf E(r}_a)+{1\over c}{\bf v}_a\crossprod{\bf B(r}_a)\right),
\end{eqnarray}
from which the constancy of $\bf P$ follows.

      Similar arguments can be carried out for a rigid rotation for which the
 change in the coordinate vector is
\def\bomega{\mbox{\boldmath{$\omega$}}}
\begin{equation}
\delta{\bf r} = \delta\bomega \crossprod{\bf r},
\end{equation}
 with $\delta\bomega$ 
 constant.  The corresponding change in a vector function is
\begin{equation}
{\bf\overline{A}(r + \delta r) = A(r) + \delta\bomega \crossprod A(r)}
\end{equation}
 since a vector transforms in the same way as $\bf r$, 
so the new function at the initial numerical values of the coordinates is
\begin{equation}
{\bf \overline{A} (r) =A (r) - (\delta r\dotprod\vecnab) A(r) + 
\delta\bomega \crossprod A(r)}.
\end{equation}
 The change in the vector potential is
\begin{equation}
{\bf \delta A = -(\delta r\dotprod \vecnab) A+ \delta\bomega \crossprod A}.
\end{equation}
 The generator can now be written in the form
\begin{equation}
 G = \delta\bomega\dotprod{\bf J},
\end{equation}
 where the total angular momentum, $\bf J$, is found to be 
\begin{equation}
{\bf J}=\sum_a{\bf r}_a\crossprod m_a{\bf v}_a+\int\deer{\bf r}
\crossprod\left({1\over4\pi c}{\bf E\crossprod B}\right),
\label{9.26}
\end{equation}
 which again is a constant of the motion.

\subsection{Gauge Invariance and the Conservation of Charge}
\label{sec9.6}
    An electromagnetic system possesses a conservation law, that of
    electric charge, which has no place in the usual mechanical framework.
    It is connected to a further invariance of the
    electromagnetic fields---the potentials
 are not uniquely defined in that if we let
\begin{equation}
{\bf A\to A }+ \vecnab\lambda,\qquad\phi\to\phi-{1\over c}{\partial
\over\partial t}\lambda,
\label{9.27}
\end{equation}
 the electric and magnetic fields defined by (\ref{9.9b}) and 
(\ref{9.10b}) remain unaltered, for an arbitrary function $\lambda$. 
This is called gauge invariance; the corresponding substitution 
(\ref{9.27}) is a gauge transformation.  [The term has its origin in
a now obsolete theory of Hermann Weyl (1885--1955) \cite{weyl1919}.] 

This invariance of
 the action must imply a corresponding conservation law.  To determine what is
 conserved, we compute the change in the Lagrangian, (\ref{9.1}), 
explicitly.  Trivially, the field part of $L$ remains unchanged.  
In considering the change of the particle part, we recognize that (\ref{9.27})
 is incomplete; since $\bf v$ is a physical quantity, 
${\bf p} - (e/c){\bf A}$
 must be invariant under a gauge transformation,
 which will only be true if (\ref{9.27}) is supplemented by
\begin{equation}
{\bf p} \to{\bf p} +{e\over c}\vecnab\lambda. 
\label{9.28}
\end{equation}
 Under the transformation (\ref{9.27}) and (\ref{9.28}), 
the Lagrangian becomes
\begin{eqnarray}
 L \to\overline{L} &\equiv& L + \sum_a\left[{e_a\over c}\vecnab\lambda
\dotprod\left({d{\bf r}_a\over dt}-{\bf v}_a\right)
+{e_a\over c} {\partial\over\partial t}\lambda+{e_a\over c}{\bf v}_a\dotprod
\vecnab\lambda\right]\nonumber\\
&=&L+\sum_a{e_a\over c}\left({\partial\over\partial t}\lambda
+{ d{\bf r}_a\over dt}\dotprod\vecnab\lambda\right)\nonumber\\
&=&L+{d\over dt}w,
\label{9.29}
\end{eqnarray}
where
\begin{equation}
w=\sum_a{e_a\over c}\lambda({\bf r}_a,t).
\label{9.30}
\end{equation}
 What is the physical consequence of adding a total time derivative to a
 Lagrangian?  It does not change the equations of motion, so the system is 
unaltered.  Since the entire change is in the end point behavior,
\begin{equation}
     \overline{W}_{12} = W_{12} +(w_1-w_2),  
\label{9.31}
\end{equation}
 the whole effect is a redefinition of the generators, $G$,
\begin{equation}
\overline{G} = G + \delta w.
\end{equation}
 This alteration reflects the fact that the Lagrangian itself is ambiguous up
 to a total time derivative term.
   
   To ascertain the implication of gauge invariance,
   we rewrite the change in the Lagrangian given in the first line 
   of (\ref{9.29}) by use of (\ref{9.5c}),
\begin{equation}
 \overline{ L}-L ={ 1\over c}\int\deer \left[\rho{\partial\over\partial t}
 \lambda+{\bf j  \dotprod\vecnab\lambda}\right],
\label{9.32}
\end{equation}
 and apply this result to an infinitesimal gauge transformation,
   $\lambda\to\delta\lambda$.
   The change in the action is then
   \begin{equation}
   \delta W_{12}=G_{\delta\lambda_1}-G_{\delta\lambda_2}-\int_{t_2}^{t_1}
   dt\int\deer{1\over c}\delta\lambda\left({\partial\over\partial t}\rho
   +\vecnab\dotprod {\bf j}\right),
   \label{1-25.13}
   \end{equation}
   with the generator being
   \begin{equation}
   G_{\delta\lambda}=\int\deer{1\over c}\rho\,\delta\lambda.
   \label{1-25.14}
   \end{equation}
   In view of the arbitrary nature of $\delta\lambda({\bf r},t)$,
   the stationary action principle now demands that, at every point,
\begin{equation}
{\partial\over\partial t}\rho+\vecnab\dotprod{\bf j}=0,
\label{9.33}
\end{equation}
that is, gauge invariance implies local charge conservation.
(Of course, this same result follows from Maxwell's equations.)
Then, the special situation $\delta\lambda=\mbox{constant}$, where
$\delta{\bf A}=\delta\phi=0$, and $W_{12}$ is certainly invariant,
implies a conservation law, that of
\begin{equation}
G_{\delta\lambda}={1\over c}\delta\lambda \,Q,
\label{1-25.16}
\end{equation}
in which 
\begin{equation}
Q=\int\deer\rho
\label{1-25.17}
\end{equation}
is the conserved total charge. 

\subsection{Gauge Invariance and Local Conservation Laws}
\label{sec9.7}
We have just derived the local conservation law of electric charge.
Electric charge is a property carried only by the particles, 
not by the electromagnetic
field.  In contrast, the mechanical properties of energy, linear momentum, and
angular momentum are attributes of both particles and fields.  For
these we have conservation laws of total quantities.  What about local
conservation laws?  The usual development of electrodynamics refers
to
local {\it non}-conservation laws; they concentrated on the fields
and characterized the charged particles as sources (or sinks) of field
mechanical properties.  It is natural to ask for a more even-handed
treatment of both charges and fields.  We shall supply it, in the framework
of a particular example.  The property of gauge invariance will be both
a valuable guide, and an aid to simplifying the calculations.

The time displacement of a complete physical system identifies its total
energy.  This suggests that time displacement of a part of the system
provides energetic information about that portion.  The ultimate limit of 
this spatial subdivision, a local description, should appear in response
to an (infinitesimal) time displacement that varies arbitrarily in space
as well as in time, $\delta t({\bf r},t)$.

Now we need a clue.  How do fields, and potentials, respond to such 
coordi\-nate-dependent displacements?  This is where the freedom of gauge
transformations enters:  The change of the vector and scalar potentials,
by $\vecnab\lambda({\bf r},t)$, $-(1/c)(\partial/\partial t)\lambda
({\bf r},t)$, respectively, serves as a model for the potentials themselves.
The advantage here is that the response of the scalar $\lambda({\bf r},t)$
to the time displacement can be reasonably taken to be
\begin{subequations}
\begin{equation}
(\lambda+\delta\lambda)({\bf r},t+\delta t)=\lambda({\bf r},t),
\label{1-26.1}
\end{equation}
or
\begin{equation}
\delta\lambda({\bf r},t)=-\delta t({\bf r},t){\partial\over\partial t}
\lambda({\bf r},t).
\label{1-26.2}
\end{equation}
\end{subequations}
Then we derive
\begin{subequations}
\begin{eqnarray}
\delta(\vecnab\lambda)&=&-\delta t{\partial\over\partial t}(\vecnab\lambda)
+\left(-{1\over c}{\partial\over\partial t}\lambda\right)c\vecnab\delta t,\\
\delta\left(-{1\over c}{\partial\over\partial t}\lambda\right)
&=&-\delta t\left(-{1\over c}{\partial^2\over\partial t^2}\lambda\right)
-\left(-{1\over c}{\partial\over\partial t}\lambda\right){\partial
\over\partial t}\delta t,
\label{1-26.3}
\end{eqnarray}
\end{subequations}
which is immediately generalized to
\begin{subequations}
\begin{eqnarray}
\delta {\bf A}&=&-\delta t{\partial\over\partial t}{\bf A}
+\phi c\vecnab\delta t,\\
\delta \phi&=&-\delta t{\partial\over\partial t}\phi-\phi{\partial
\over\partial t}\delta t,
\label{1-26.4}
\end{eqnarray}
\end{subequations}
or, equivalently,
\begin{subequations}
\begin{eqnarray}
\delta {\bf A}&=&c\delta t{\bf E}+\vecnab(\phi c\delta t),\\
\delta\phi&=&-{1\over c}{\partial\over\partial t}(\phi c\delta t).
\label{1-26.5}
\end{eqnarray}
\end{subequations}
In the latter form we recognize a gauge transformation, produced by the
scalar $\phi c\delta t$, which will not contribute to the changes of
field strengths.  Accordingly, for that calculation we have,
effectively, $\delta{\bf A}=c\delta t{\bf E}$, $\delta \phi=0$, leading
to
\begin{subequations}
\begin{eqnarray}
\delta{\bf E}&=&-{1\over c}{\partial\over\partial t}(c\delta t {\bf E})
=-\delta t{\partial\over\partial t}{\bf E}-{\bf E}{\partial\over
\partial t}\delta t,\\
\delta{\bf B}&=&\vecnab\crossprod(c\delta t{\bf E})=-\delta t
{\partial\over\partial t}{\bf B}-{\bf E}\crossprod\vecnab
c\delta t;
\label{1-26.6}
\end{eqnarray}
\end{subequations}
the last line employs the field equation $\vecnab\crossprod{\bf E}
=-(1/c)(\partial{\bf B}/\partial t)$.

In the following we adopt a viewpoint in which such homogeneous field
equations are accepted as consequences of the definition of the fields
in terms of potentials.  That permits the field Lagrange function
(\ref{9.1}) to be simplified:
\begin{equation}
{\cal L}_{\rm field}={1\over8\pi}(E^2-B^2).
\label{1-26.7}
\end{equation}
Then we can apply the field variation (\ref{1-26.6}) directly, and get
\begin{eqnarray}
\delta{\cal L}_{\rm field}&=&-\delta t{\partial\over\partial t}
{\cal L}_{\rm field}-{1\over4\pi}E^2{\partial\over\partial t}
\delta t-{c\over4\pi}{\bf E\crossprod B}\dotprod\vecnab\delta t\nonumber\\
&=&-{\partial\over\partial t}(\delta t{\cal L}_{\rm field})
-{1\over8\pi}(E^2+B^2){\partial\over\partial t}\delta t
-{c\over4\pi}{\bf E\crossprod B}\dotprod\vecnab\delta t.
\label{1-26.8}
\end{eqnarray}
Before commenting on these last, not unfamiliar, field structures, we turn
to the charged particles and put them on a somewhat similar footing
in terms of a continuous, rather than a discrete, description.

We therefore present the Lagrangian of the charges in (\ref{9.1}) in
terms of a corresponding Lagrange function,
\begin{subequations}
\begin{equation}
L_{\rm charges}=\int\deer{\cal L}_{\rm charges},
\label{1-26.9}
\end{equation}
where
\begin{equation}
{\cal L}_{\rm charges}=\sum_a{\cal L}_a
\label{1-26.10}
\end{equation}
and
\begin{equation}
{\cal L}_a=\delta({\bf r-r}_a(t))\left[{1\over2}m_av_a(t)^2
-e_a\phi({\bf r}_a,t)+{e_a\over c}{\bf v}_a(t)\dotprod{\bf A(r}_a,t)
\right];
\label{1-26.11}
\end{equation}
\end{subequations}
the latter adopts the Lagrangian viewpoint, with ${\bf v}_a=d{\bf r}_a/dt$
accepted as a definition.  Then, the effect of the time displacement
on the variables ${\bf r}_a(t)$, taken as
\begin{subequations}
\begin{eqnarray}
({\bf r}_a+\delta{\bf r}_a)(t+\delta t)&=&{\bf r}_a(t),\label{1-26.12}\\
\delta{\bf r}_a(t)&=&-\delta t({\bf r}_a,t){\bf v}_a(t),\label{1-26.13}
\end{eqnarray}
\end{subequations}
implies the velocity variation
\begin{equation}
\delta{\bf v}_a(t)=-\delta t({\bf r}_a,t){d\over dt}{\bf v}_a(t)
-{\bf v}_a(t)\left[{\partial\over\partial t}\delta t+{\bf v}_a\dotprod
\vecnab\delta t\right];
\label{1-26.14}
\end{equation}
the last step exhibits both the explicit and the implicit dependences
of $\delta t({\bf r}_a,t)$ on $t$.  In computing the variation of 
$\phi({\bf r}_a,t)$, for example, we combine the potential variation
given in (\ref{1-26.4}) with the effect of $\delta{\bf r}_a$:
\begin{subequations}
\begin{equation}
\delta\phi({\bf r}_a(t),t)=-\delta t{\partial\over\partial t}\phi
-\phi{\partial\over\partial t}\delta t-\delta t{\bf v}_a\dotprod
\vecnab_a\phi=-\delta t{d\over dt}\phi-\phi{\partial\over\partial t}
\delta t,
\label{1-26.15}
\end{equation}
and, similarly,
\begin{equation}
\delta{\bf A}({\bf r}_a(t),t)=-\delta t{\partial\over\partial t}{\bf A}
+\phi c\vecnab\delta t-\delta t{\bf v}_a\dotprod
\vecnab_a{\bf A}=-\delta t{d\over dt}{\bf A}+\phi c
\vecnab\delta t.
\label{1-26.16}
\end{equation}
\end{subequations}

The total effect of these variations on ${\cal L}_a$ is thus
\begin{subequations}
\begin{equation}
\delta{\cal L}_a=-\delta t{d\over dt}{\cal L}_a+\delta({\bf r-r}_a(t))
\left(-m_av_a^2-{e_a\over c}{\bf A\dotprod v}_a+e_a\phi\right)
\left({\partial\over\partial t}\delta t+{\bf v}_a\dotprod\vecnab\delta t
\right),
\label{1-26.17}
\end{equation}
or
\begin{equation}
\delta{\cal L}_a=-{d\over dt}(\delta t{\cal L}_a)-\delta({\bf r-r}_a(t))
E_a\left({\partial\over\partial t}\delta t+{\bf v}_a\dotprod\vecnab
\delta t\right),
\label{1-26.18}
\end{equation}
\end{subequations}
where we see the kinetic energy of the charged particle,
\begin{equation}
E_a={1\over2}m_av_a^2.
\label{1-26.19}
\end{equation}
We have retained the particle symbol $d/dt$ to the last, but now, 
being firmly back in the field, space-time viewpoint, it should be
written as $\partial/\partial t$, referring to all $t$ dependence,
with $\bf r$ being held fixed.  The union of these various 
contributions to the variation of the total Lagrange function is
\begin{equation}
\delta{\cal L}_{\rm tot}=-{\partial\over\partial t}(\delta t
{\cal L}_{\rm tot})-U_{\rm tot}{\partial\over\partial t}\delta t
-{\bf S}_{\rm tot}\dotprod\vecnab\delta t,
\label{1-26.20}
\end{equation}
where, from (\ref{1-26.8}) and (\ref{1-26.18}),
\begin{subequations}
\begin{equation}
U_{\rm tot}={1\over8\pi}(E^2+B^2)+\sum_a\delta({\bf r-r}_a(t))E_a
\label{1-26.21}
\end{equation}
and
\begin{equation}
{\bf S}_{\rm tot}={c\over4\pi}{\bf E\crossprod B}+\sum_a\delta
({\bf r-r}_a(t))E_a{\bf v}_a,
\label{1-26.22}
\end{equation}
\end{subequations}
are physically transparent forms for the total energy density and total
energy flux vector.

To focus on what is new in this development, we ignore boundary effects
in the stationary action principle, by setting the otherwise arbitrary
$\delta t({\bf r},t)$ equal to zero at $t_1$ and $t_2$.  Then, through
partial integration, we conclude that
\begin{equation}
\delta W_{12}=\int_{t_2}^{t_1}dt\int\deer\delta t\left({\partial\over
\partial t}U_{\rm tot}+\vecnab\dotprod{\bf S}_{\rm tot}\right)=0,
\label{1-26.23}
\end{equation}
from which follows the local statement of total energy conservation,
\begin{equation}
{\partial\over\partial t}U_{\rm tot}+\vecnab\dotprod{\bf S}_{\rm tot}=0,
\label{1-26.24}.
\end{equation}








%% file: chap2.tex
This section, and the following three, are based on lectures given by
the author in quantum field theory courses at the University of Oklahoma
over several years, based in turn largely on lectures given by Schwinger at
Harvard in the late 1960s.

After the above reminder of classical variational principles,
we now turn to the dynamics of quantum mechanics.  We begin by considering
the transformation function $\langle a',t+dt|b',t\rangle$.  Here
$|b',t\rangle$ is a state specified by the values $b'=\{b'\}$ of a
complete set of dynamical variables $B(t)$, while $|a',t+dt\rangle$
is a state specified by values  $a'=\{a'\}$ of a (different) complete set
of dynamical variables $A(t+dt)$, defined at a slightly later time.\footnote{ 
Here Schwinger is using his standard notation, designating eigenvalues by 
primes.} 
We suppose that $A$ and $B$ do not possess
any explicit time dependence---that is, their definition does not depend
upon $t$.  Here
\be
\langle a',t+dt|=\langle a',t|U,
\ee
where the infinitesimal time translation operator is related to the generator
of time translations as follows,
\be
U=1+iG=1-i\,dt\,H.
\ee
The Hamiltonian $H$ is a function of dynamical variables, which we write 
generically as $\chi(t)$, and of $t$ explicitly.  Thus
\be
\langle a',t+dt|b',t\rangle=\langle a',t|1-i\,dt\,H(\chi(t),t)|b',t\rangle.
\ee

We next translate states and operators to time zero:
\begin{subequations}
\bea
\langle a',t|&=&\langle a'|U(t),\quad |b',t\rangle=U^{-1}(t)|b'\rangle,\\
\chi(t)&=&U^{-1}(t)\chi U(t),
\eea
\end{subequations}
where $\chi=\chi(0)$, etc.  Then,
\be
\langle a',t+dt|b',t\rangle=\langle a'|1-i\,dt\,H(\chi,t)|b'\rangle,\ee
or, as a differential equation
\bea
\delta_{\rm dyn}\langle a',t+dt|b',t\rangle&=&i\langle a'|\delta_{\rm dyn}[-dt H]
|b'\rangle\nonumber\\
&=&i\langle a',t+dt|\delta_{\rm dyn}[-dt\,H(\chi(t),t)]|b',t\rangle,
\label{dynvar}
\eea
where $\delta_{\rm dyn}$ corresponds to changes in initial and final times,
$\delta t_2$ and $\delta t_1$, and in the structure of $H$, $\delta H$.
[By reintroducing $dt$ in the state on the left in 
the second line, we make a negligible error of ${\cal O}(dt^2)$.]

However, we can also consider {\it kinematical\/} changes.  To understand
these, consider a system defined by coordinates and momenta, $\{q_a(t)\}$,
$\{p_a(t)\}$, $a=1,\dots,n$, which satisfy the canonical commutation relations,
\begin{subequations}
\bea[q_a(t),p_b(t)]&=&i\delta_{ab},\quad(\hbar=1)\\
\mbox{}[q_a(t),q_b(t)]&=&[p_a(t),p_b(t)]=0.
\eea
\end{subequations}
A spatial displacement $\delta q_a$ is induced by
\be
U=1+iG_q,\quad G_q=\sum_{a=1}^n p_a\delta q_a.
\ee
In fact ($\delta q_a$ is a number, not an operator), 
\bea
U^{-1}q_aU&=&q_a-\frac1i[q_a,G_q]\nonumber\\
&=&q_a-\delta q_a,
\eea
while
\be
U^{-1}p_aU=p_a-\frac1i[p_a,G_q]=p_a.
\ee

The (dual) symmetry between position and momentum,
\be
q\to p,\quad p\to-q,
\ee
gives us the form for the generator of a displacement in $p$:
\be
G_p=-\sum_aq_a\delta p_a.
\ee

A {\it kinematic\/} variation in the states is given by the generators
\begin{subequations}
\bea
\delta_{\rm kin}\langle\,\,\,|&=&\overline{\langle\,\,\,|}-\langle\,\,\,|
=\langle\,\,\,|iG,\\
\delta_{\rm kin}|\,\,\,\rangle&=&\overline{|\,\,\,\rangle}-|\,\,\,\rangle
=-iG|\,\,\,\rangle,
\eea\end{subequations}
so, for example, under a $\delta q$ variation, the transformation function
changes by
\be
\delta_q\langle a',t+dt|b',t\rangle=i\langle a',t+dt|\sum_a\left[p_a(t+dt)
\delta q_a(t+dt)-p_a(t)\delta q_a(t)\right]|b',t\rangle.
\ee
Now the dynamical variables at different times are related by Hamilton's 
equations,
\bea
\frac{d p_a(t)}{dt}&=&\frac1i[p_a(t),H(q(t),p(t),t)]\nonumber\\
&=&-\frac{\partial H}{\partial q_a}(t),
\eea
so
\be
p_a(t+dt)-p_a(t)=dt\frac{dp_a(t)}{dt}=-dt\frac{\partial H}{\partial q_a}(t).
\ee
Similarly, the other Hamilton's equation
\be
\frac{dq_a}{dt}=\frac{\partial H}{\partial p_a}
\ee
implies that
\be
q_a(t+dt)-q_a(t)=dt\frac{\partial H}{\partial p_a}(t).
\ee
From this we deduce first the $q$ variation of the transformation function,
\bea
&&\delta_q\langle a',t+dt|b',t\rangle\nonumber\\
&=&i\langle a',t+dt|\sum_ap_a(t)[\delta
q_a(t+dt)-\delta q_a(t)]-dt\frac{\partial H}{\partial q_a}\delta q_a(t)
+{\cal O}(dt^2)|b',t\rangle\nonumber\\
&=&i\langle a',t+dt|\delta_q\left[\sum_ap_a(t)\mbox{.}[q_a(t+dt)-q_a(t)]
-dt\,H(q(t),p(t),t)\right]|b',t\rangle,\nonumber\\
\eea
where the dot denotes symmetric multiplication of the $p$ and $q$ operators.

For $p$ variations we have a similar result:
\bea
&&\delta_p\langle a',t+dt|b',t\rangle\nonumber\\
&=&-i\langle a',t+dt|\sum_a[q_a(t+dt)
\delta p_a(t+dt)-q_a(t)\delta p_a(t)]|b',t\rangle\nonumber\\
&=&-i\langle a',t+dt|\sum_a q_a(t)[\delta p_a(t+dt)-\delta p_a(t)]+dt\,
\frac{\partial H}{\partial p_a}(t)\delta p_a(t)|b',t\rangle\nonumber\\
&=&i\langle a',t+dt|\delta_p\left[-\sum_aq_a(t)\mbox{.}(p_a(t+dt)-p_a(t))-dt\,
H(q(t),p(t),t)\right]|b',t\rangle.\nonumber\\
\eea
That is, for $q$ variations
\begin{subequations}
\be
\delta_q\langle a',t+dt|b',t\rangle=i\langle a',t+dt|\delta_q\left[dt L_q
\right]|b',t\rangle,
\ee
with the quantum Lagrangian
\be
L_q=\sum_ap_a\mbox{.}\dot q_a-H(q,p,t),
\ee
\end{subequations}
while for $p$ variations
\begin{subequations}
\be
\delta_p\langle a',t+dt|b',t\rangle=i\langle a',t+dt|\delta_p\left[dt L_p
\right]|b',t\rangle,
\ee
with the quantum Lagrangian
\be
L_p=-\sum_aq_a\mbox{.}\dot p_a-H(q,p,t).
\ee
\end{subequations}
We see here two alternative forms of the quantum Lagrangian.  Note that the
two forms differ by a total time derivative,
\be
L_q-L_p=\frac{d}{dt}\sum_a p_a\mbox{.} q_a.
\ee

We now can unite the kinematic transformations considered here with the 
dynamic ones considered earlier, in Eq.~(\ref{dynvar}):
\be
\delta=\delta_{\rm dyn}+\delta_{\rm kin}:\quad
\delta\langle a',t+dt|b',dt\rangle=i\langle a',t+dt|\delta[dt\,L]|b',t\rangle.
\label{infvp}
\ee

Suppose, for concreteness, that our states are defined by values of $q$, so
that
\be
\delta_p\langle a',t+dt|b't\rangle=0.
\ee
This is consistent, as a result of Hamilton's equations,
\be
\delta_pL_q=\sum_a\delta p_a\left(\dot q_a-\frac{\partial H}{\partial p_a}
\right)=0.
\ee
In the following we will use $L_q$.

It is immediately clear that we can iterate the infinitesimal version
(\ref{infvp})
of the quantum action principle by inserting at each time step a complete
set of intermediate states (to simplify the notation, we ignore their quantum
numbers):
\be
\langle t_1|t_2\rangle=\langle t_1|t_1-dt\rangle\langle t_1-dt|t_1-2dt\rangle
\cdots\langle t_2+2dt|t_2+dt\rangle\langle t_2+dt|t_2\rangle,
\ee
So in this way we deduce the general form of {\em Schwinger's quantum
action principle:}
\be
\delta\langle t_1|t_2\rangle=i\langle t_1|\delta\int_{t_2}^{t_1}dt\, L
|t_2\rangle.
\label{qap}
\ee
This summarizes all the properties of the system.

Suppose the dynamical system is given, that is, the structure of $H$ does not
change.  Then
\be
\delta\langle t_1|t_2\rangle=i\langle t_1|G_1-G_2|t_2\rangle,
\ee
where the generator $G_a$ depends on $p$ and $q$ at time $t_a$.  Comparing
with the action principle (\ref{qap}) we see
\be
\delta\int_{t_2}^{t_1}dt\,L=G_1-G_2,
\ee
which has exactly the form of the classical action principle 
(\ref{8.2}), except that
the Lagrangian $L$ and the generators $G$ are now operators.
If no changes occur at the endpoints, we have the {\it principle of
stationary action},
\be
\delta\int_{t_2}^{t_1}\left(\sum_a p_a.dq_a-H\,dt\right)=0.
\ee
As in the classical case, let us introduce a time parameter $\tau$, 
$t=t(\tau)$, such that $\tau_2$ and $\tau_1$ are fixed.  The the above
variation reads
\bea
&&\sum_a\left[\delta p_a.d q_a+p_a.d\delta q_a-\delta H\,dt-H\,d\delta t\right]
\nonumber\\
&=&d\left[\sum_ap_a.\delta q_a-H\,\delta t\right]
+\sum_a\left[\delta p_a.d q_a-dp_a.\delta q_a\right]-\delta H\,dt+dH\,\delta t,
\eea
so the action principle says
\begin{subequations}
\bea
G&=&\sum_ap_a.\delta q_a-H\,\delta t,\\
\delta H&=&\frac{dH}{dt}\delta t+\sum_a\left(\delta p_a.\frac{dq_a}{dt}
-\delta q_a.\frac{dp_a}{dt}\right).
\eea
\end{subequations}
We will again assume $\delta p_a$, $\delta q_a$ are not operators (that is,
they are proportional to the unit operator); then we recover Hamilton's
equations,
\begin{subequations}
\bea
\frac{\partial H}{\partial t}&=&\frac{dH}{dt},\\
\frac{\partial H}{\partial p_a}&=&\frac{dq_a}{dt},\\
\frac{\partial H}{\partial q_a}&=&-\frac{dp_a}{dt}.
\eea
\end{subequations}
(Schwinger also explored the possibility of operator variations
\cite{leshouches}.)
We learn from the generators,
\be
G_t=-H\,\delta t,\quad G_q=\sum_ap_a\delta q_a,
\ee
that the change in some function $F$ of the dynamical variable is
\be
\delta F=\frac{dF}{dt}\delta t+\frac1i[F,G],
\ee
so we deduce
\begin{subequations}
\bea
\frac{dF}{dt}&=&\frac{\partial F}{\partial t}+\frac1i[F,H],\\
\frac{\partial F}{\partial q_a}&=&\frac1i[F,p_a].
\eea
\end{subequations}
Note that from this the canonical commutation relations follow,
\be
[q_a,p_b]=i\delta_{ab},\quad[p_a,p_b]=0,
\ee
as well as Newton's law,
\be
\dot p_a=-\frac1i[H,p_a]=-\frac{\partial H}{\partial q_a}.
\ee

If we had used $L_p$ instead of $L_q$, we would have obtained the same
equations of motion, but in place of $G_q$, we would have obtained
\be
G_p=-\sum_aq_a\delta p_a,
\ee
which implies
\be
\frac{\partial F}{\partial p_a}=-\frac1i[F,q_a].
\ee
From this can be deduced the remaining canonical commutator,
\be
[q_a,q_b]=0,
\ee
as well as the remaining Hamilton equation,
\be
\dot q_a=\frac1i[q_a,H]=\frac{\partial H}{\partial p_a}.
\ee
It is easy to show that the effect of changing the Lagrangian
by a total time derivative (which is what is done in passing from $L_q$
to $L_p$) is to change the generators.

We now turn to examples.

\section{Harmonic Oscillator}
The harmonic oscillator is defined in terms of creation and annihilation
operators,\footnote{We follow Schwinger's usage of $y$ for the annihilation
operator, instead of the more usual $a$.}
 $y^\dagger$ and $y$, and the corresponding Hamiltonian $H$,
\begin{subequations}
\bea
[y,y^\dagger]&=&1,\\
H&=&\omega\left(y^\dagger y+\frac12\right).
\eea
\end{subequations}
The equations of motion are
\begin{subequations}
\label{hoeom}
\bea
\frac{dy}{dt}&=&\frac1i[y,H]=\frac1i\omega y,\label{hoeoma}\\
\frac{dy^\dagger}{dt}&=&\frac1i[y^\dagger,H]=-\frac1i\omega y^\dagger.
\label{hoeomb}
\eea
\end{subequations}
Eigenstates of $y$ and $y^\dagger$ exist, as right and left vectors,
respectively,
\begin{subequations}
\bea
y|y'\rangle&=&y'|y'\rangle,\\
\langle y^{\dagger\prime}|y^\dagger&=&y^{\dagger\prime}\langle 
y^{\dagger\prime}|,
\eea
\end{subequations}
while $\langle y'|$ and $|y^{\dagger\prime}\rangle$ do not exist.\footnote{If
$\langle y'|y=y'\langle y'|$ then we would have an evident contradiction:
\be
1=\langle y'|[y,y^\dagger]|y'\rangle=y'\langle y'|y^\dagger|y'\rangle
-\langle y'|y^\dagger|y'\rangle y'=0.
\ee}  These are the famous ``coherent states,'' to whom the name
Roy Glauber \cite{glauber} is invaribly attached, although they were discovered
by Erwin Schr\"odinger \cite{Schrodinger1926}, and Glauber's approach, as he 
acknowledged, follewed that of his mentor, Schwinger \cite{Schwinger1953}.

The transformation function we seek is therefore
\be
\langle y^{\dagger\prime},t_1|y'',t_2\rangle.
\ee
If we regard $y$ as a ``coordinate,'' the corresponding ``momentum'' is
$iy^\dagger$:
\be
\dot y=\frac1i\omega y=\frac{\partial H}{\partial iy^\dagger},\quad
i\dot y^\dagger=-\omega y^\dagger=-\frac{\partial H}{\partial y}.
\ee
The corresponding Lagrangian is therefore\footnote{We might note that
in terms of (dimensionless) position and momentum operators
\be
i y^\dagger.\dot y=\frac{i}2(q-ip).(\dot q+i\dot p)=\frac12(p.\dot q
-q.\dot p)+\frac{i}4\frac{d}{dt}(q^2+p^2),
\ee
where the first term in the final form is the average of the Legendre
transforms in $L_q$ and $L_p$.}
\be
L=iy^\dagger.\dot y-H.
\ee
Because we use $y$ as our state variable at the initial time, and $y^\dagger$
at the final time, we must exploit our freedom to redefine our generators
to write
\be
W_{12}=\int_2^1dt\,L-iy^\dagger(t_1).y(t_1).
\ee
Then the variation of the action is
\bea
\delta W_{12}&=&-i\delta(y_1^\dagger.y_1)+G_1-G_2\nonumber\\
&=&-i\delta y^\dagger_1.y_1-iy^\dagger_1.\delta y_1+iy^\dagger_1.\delta y_1
-iy^\dagger_2.\delta y_2-H\,\delta t_1+H\,\delta t_2\nonumber\\
&=&-i\delta y^\dagger_1.y_1-iy_2^\dagger.\delta y_2-H(\delta t_1-\delta t_2).
\eea
Then the quantum action principle says
\be
\delta\langle y^{\dagger\prime},t_1|y'',t_2\rangle=i\langle y^{\dagger\prime}
,t_1|-i\delta y_1^{\dagger\prime}y_1-iy_2^\dagger\delta y_2''
-\omega y_1^{\dagger\prime}y_1(\delta t_1-\delta t_2)|y'',t_2\rangle,
\ee
since by assumption the variations in the dynamical variables are numerical:
\be
[\delta y_1^\dagger,y_1]=[y_2^\dagger,\delta y_2],
\ee
and we have dropped the zero-point energy.
Now use the equations of motion (\ref{hoeoma}) and (\ref{hoeomb}) 
to deduce that
\be
y_1=e^{-i\omega(t_1-t_2)}y_2,\quad y_2^\dagger=e^{-i\omega(t_1-t_2)}y_1^\dagger
\ee
and hence
\bea
\delta\langle y^{\dagger\prime},t_1|y'',t_2\rangle
&=&\langle y^{\dagger\prime},t_1|\delta y^{\dagger\prime}e^{-i\omega(t_1-t_2)}
y''+y^{\dagger\prime}e^{-i\omega(t_1-t_2)}\delta y''\nonumber\\
&&\quad\mbox{}-i\omega y^{\dagger\prime}
e^{-i\omega(t_1-t_2)}(\delta t_1-\delta t_2)y''|y'',t_2\rangle\nonumber\\
&=&\langle y^{\dagger\prime},t_1|y'',t_2\rangle\delta\left[y^{\dagger\prime}
e^{-i\omega(t_1-t_2)}y''\right].
\eea
From this we can deduce that the transformation function has the exponential
form
\be
\langle y^{\dagger\prime},t_1|y'',t_2\rangle=\exp\left[y^{\dagger\prime}
e^{-i\omega(t_1-t_2)}y''\right],
\label{hosoln}
\ee
which has the correct boundary condition at $t_1=t_2$; and in particular,
$\langle 0|0\rangle=1$.

On the other hand, 
\be
\langle y^{\dagger\prime},t_1|y'',t_2\rangle=\langle y^{\dagger\prime}|
e^{-iH(t_1-t_2)}|y''\rangle,
\ee
where both states are expressed at the common time $t_2$, so, upon
inserting a complete set of energy eigenstates, we obtain ($t=t_1-t_2$)
\be
\sum_E\langle y^{\dagger\prime}|E\rangle e^{-iEt}\langle E|y''\rangle,
\ee
which we compare to the Taylor expansion of the previous formula,
\be
\sum_{n=0}^\infty \frac{(y^{\dagger\prime})^n}{\sqrt{n!}}e^{-in\omega t}
\frac{(y'')^n}{\sqrt{n!}}.
\ee
This gives all the eigenvectors and eigenvalues:
\begin{subequations}
\bea
E_n&=&n\omega,\quad n=0,1,2,\dots,\\
\langle y^{\dagger\prime}|E_n\rangle&=&\frac{(y^{\dagger\prime})^n}{\sqrt{n!}},
\label{cohstton}\\
\langle E_n|y''\rangle&=&\frac{(y'')^n}{\sqrt{n!}}.
\eea
\end{subequations}
These correspond to the usual construction of the eigenstates from the
ground state:
\be
|E_n\rangle=\frac{(y^\dagger)^n}{\sqrt{n!}}|0\rangle.
\ee

\section{Forced Harmonic Oscillator}
\label{sec:chap2}
Now we add a driving term to the Hamiltonian,
\be
H=\omega y^\dagger y+yK^*(t)+y^\dagger K(t),\label{forcedho}
\ee
where $K(t)$ is an external force ({\it Kraft\/} is force in German).
The equation of motion is
\be
i\frac{dy}{dt}=\frac{\partial H}{\partial y^\dagger}=[y,H]=\omega y+K(t),
\label{eom:fho}
\ee
while $y^\dagger$ satisfies the adjoint equation.  In the presence of $K(t)$,
we wish to compute the transformation function
$\langle y^{\dagger\prime},t_1|y'',t_2\rangle^K$.

Consider a variation of $K$.  According to the action principle
\bea
\delta_K\langle y^{\dagger\prime},t_1|y'',t_2\rangle^K&=&
\langle y^{\dagger\prime},t_1|i\delta_KW_{12}|y'',t_2\rangle^K
\nonumber\\
&=&-i\langle y^{\dagger\prime},t_1|\int_{t_2}^{t_1}dt[\delta K y^\dagger
+\delta K^* y]|y'',t_2\rangle^K.
\label{de:fho}
\eea
We can solve this differential equation by noting that the equation of
motion (\ref{eom:fho}) can be rewritten as
\be
i\frac{d}{dt}\left[e^{i\omega t}y(t)\right]=e^{i\omega t}K(t),
\ee
which is integrated to read
\be
e^{i\omega t}y(t)-e^{i\omega t_2}y(t_2)=-i\int_{t_2}^t dt'\,e^{i\omega t'}
K(t'),
\ee
or
\be
y(t)=e^{-i\omega(t-t_2)}y_2-i\int_{t_2}^t dt'\,e^{-i\omega(t-t')}K(t'),
\label{first}
\ee
and the adjoint\footnote{The consistency of these two equations
follows from
\be
e^{i\omega t_1}y_1=e^{i\omega t_2}y_2-i\int_{t_2}^{t_1}dt'\,e^{i\omega t'}
K(t'),\ee
so that the adjoint of Eq.~(\ref{first}) is
\bea
[y(t)]^\dagger&=&e^{i\omega t}\left[e^{-i\omega t_1}y_1^\dagger-i\int_{t_2}
^{t_1}dt'\,e^{-i\omega t'}K^*(t')\right]+i\int_{t_2}^tdt'\,e^{-i\omega(t'-t)}
K^*(t')\nonumber\\
&=&e^{i\omega(t-t_1)}y_1^\dagger+i\int_{t_1}^tdt'\,e^{-i\omega(t'-t)}K^*(t'),
\eea
which is Eq.~(\ref{second}).}
\be
y^\dagger(t)=e^{-i\omega(t_1-t)}y^\dagger_1
-i\int_{t}^{t_1} dt'\,e^{-i\omega(t'-t)}K^*(t').
\label{second}
\ee
Thus our differential equation (\ref{de:fho}) reads
\bea
&&\frac{\delta_K\langle y^{\dagger\prime},t_1|y'',t_2\rangle^K}
{\langle y^{\dagger\prime},t_1|y'',t_2\rangle^K}=
\delta_K\ln\langle y^{\dagger\prime},t_1|y'',t_2\rangle^K
\nonumber\\
&&\qquad=-i\int_{t_2}^{t_1}dt\,\delta K(t)\left[y^{\dagger\prime}
e^{-i\omega(t_1-t)}
-i\int_t^{t_1}dt'\,e^{-i\omega(t'-t)}K^*(t')\right]\nonumber\\
&&\qquad\quad\mbox{}-i\int_{t_2}^{t_1}dt\,\delta K^*(t)
\left[e^{-i\omega(t-t_2)}y''
-i\int_{t_2}^{t}dt'\,e^{-i\omega(t-t')}K(t')\right].
\eea
Notice that in the terms bilinear in $K$ and $K^*$, $K$ always occurs 
earlier than $K^*$.  Therefore, these terms can be combined to read
\be
-\delta_K\int_{t_2}^{t_1}dt\,dt'\,K^*(t)\eta(t-t')e^{-i\omega(t-t')}K(t'),
\ee
where the step function is
\be
\eta(t)=\left\{\begin{array}{cc}
1,&t>0,\\
0,&t<0.
\end{array}\right.
\ee
Since we already know the $K=0$ value from Eq.~(\ref{hosoln}),
we may now immediately integrate our differential equation:
\bea
\langle y^{\dagger\prime},t_1|y'',t_2\rangle^K&=&\exp\bigg[y^{\dagger\prime}
e^{-i\omega(t_1-t_2)}y''\nonumber\\
&&\quad\mbox{}-iy^{\dagger\prime}\int_{t_2}^{t_1}dt\,e^{-i\omega(t_1-t)}K(t)
-i\int_{t_2}^{t_1}dt \,e^{-i\omega(t-t_2)}K^*(t)\,y''\nonumber\\
&&\quad\mbox{}-\int_{t_2}^{t_1}dt\,dt'\,K^*(t)\eta(t-t')e^{-i\omega(t-t')}
K(t')\bigg].
\label{fhosln}
\eea
The ground state is defined by $y''=y^{\dagger\prime}=0$, so
\be
\langle 0,t_1|0,t_2\rangle^K=\exp\left[-\int_{-\infty}^\infty dt\,dt'\,
K^*(t)\eta(t-t')e^{-i\omega(t-t')}K(t')\right],
\label{vpax}
\ee
where we now suppose that the forces turn off at the initial and final times,
$t_2$ and $t_1$, respectively.

A check of this result is obtained by computing the probability of the system
remaining in the ground state:
\bea
|\langle 0,t_1|0,t_2\rangle^K|^2&=&\exp\bigg\{-\int_{-\infty}^\infty dt\,dt'\,
K^*(t)e^{-i\omega(t-t')}[\eta(t-t')+\eta(t'-t)]K(t')\bigg\}\nonumber\\
&=&\exp\left[-\int_{-\infty}^\infty
 dt\,dt'\,K^*(t)e^{-i\omega(t-t')}K(t')\right]\nonumber\\
&=&\exp\left[-|K(\omega)|^2\right],
\eea
where the Fourier transform of the force is
\be
K(\omega)=\int_{-\infty}^\infty dt\,e^{i\omega t}K(t).
\ee
The probability requirement
\be
|\langle 0,t_1|0,t_2\rangle^K|^2\le1
\ee
is thus satisfied.  We see here a {\em resonance\/} effect: 
If the oscillator is
driven close to its natural frequency, so $K(\omega)$ is
large, there is a large probability of finding the system in an excited
state, and therefore of not remaining in the ground state.
Let us calculate this transition amplitude to an excited state.  By
setting $y''=0$ in Eq.~(\ref{fhosln}) we obtain
\bea
\langle y^{\dagger\prime},t_1|0,t_2\rangle^K&=&\exp\left[-iy^{\dagger\prime}
\int_{-\infty}^\infty dt\,e^{-i\omega(t_1-t)}K(t)\right]\langle 0,t_1|
0,t_2\rangle^K\nonumber\\
&=&\sum_n \langle y^{\dagger\prime},t_1|n,t_1\rangle\langle n,t_1|0,t_2
\rangle^K,
\label{vactocoh}
\eea
where we have inserted a sum over a complete set of energy eigenstates,
which possess the amplitude [see Eq.~(\ref{cohstton})]
\be
\langle y^{\dagger\prime}|n\rangle=\frac{(y^{\dagger\prime})^n}{\sqrt{n!}}.
\ee
If we expand the first line of Eq.~(\ref{vactocoh}) in powers of
 $y^{\dagger\prime}$, we find
\be
\langle n,t_1|0,t_2\rangle^K=\frac{(-i)^n}{\sqrt{n!}}e^{-in\omega t_1}
[K(\omega)]^n\langle0,t_1|0,t_2\rangle^K.
\ee
The corresponding probability is
\be
p(n,0)^K=|\langle n,t_1|0,t_2\rangle^K|^2=\frac{|K(\omega)|^{2n}}{n!}
e^{-|K(\omega)|^2},\label{pnok}
\ee
which is a Poisson distribution\footnote{A Poisson probability distribution
has the form
$p(n)=\lambda^ne^{-\lambda}/n!$. The mean value of $n$ for this distribution
is
\bea
\bar n&=&\sum_{n=0}^\infty n\,p(n)=\sum_{n=0}^\infty\frac{\lambda^n 
e^{-\lambda}}{(n-1)!}=\lambda\sum_{n=0}^\infty p(n)=\lambda.
\eea}
with mean $\bar n=|K(\omega)|^2$.

Finally, let us define the {\it Green's function\/} for this problem by
\be
G(t-t')=-i\eta(t-t')e^{-i\omega(t-t')}.
\label{hogf}
\ee
It satisfies the differential equation
\be
\left(i\frac{d}{dt}-\omega\right)G(t-t')=\delta(t-t'),\label{diffeq:gf}
\ee
as it must because [see Eq.~(\ref{eom:fho})]
\be
\left(i\frac{d}{dt}-\omega\right)y(t)=K(t),
\ee
where $y(t)$ is given by [see Eq.~(\ref{first})]
\be
y(t)=e^{-i\omega(t-t_2)}y_2+\int_{-\infty}^\infty dt'\,G(t-t')K(t').
\label{yoft}
\ee
Similarly, from Eq.~(\ref{second})
\be
y^\dagger(t)=e^{-i\omega(t_1-t)}y_1^\dagger
+\int_{-\infty}^\infty dt'\,G(t'-t)K^*(t').
\ee
We can now write the ground-state persistence amplitude (\ref{vpa}) as
\be
\langle 0,t_1|0,t_2\rangle^K=\exp\left[-i\int_{-\infty}^\infty dt\,dt'\,
K^*(t)G(t-t')K(t')\right],
\label{gspa}
\ee
and the general amplitude (\ref{fhosln}) as
\bea
\langle y^{\dagger\prime},t_1|y'',t_2\rangle^K&=&
\exp\bigg\{-i\int_{-\infty}^\infty dt\,dt'\left[K^*(t)+iy^{\dagger\prime}
\delta(t-t_1)\right]\nonumber\\
&&\quad\times G(t-t')\left[K(t')+iy''\delta(t'-t_2)\right]\bigg\},
\eea
which demonstrates that knowledge of $\langle 0,t_1|0,t_2\rangle^K$
for all $K$ determines everything:
\be
\langle y^{\dagger\prime},t_1|y'',t_2\rangle^K=\langle 0,t_1|0,t_2
\rangle^{K(t)+iy''\delta(t-t_2)+iy^{\dagger\prime}\delta(t-t_1)}.
\ee

\section{Feynman Path Integral Formulation}

Although much more familiar, the path integral formulation of quantum
mechanics \cite{feynman,feynmanst,Feynman1965}
 is rather vaguely defined.  We will here provide a formal 
``derivation'' based on the Schwinger principle, in the harmonic oscillator
context.  

Consider a forced oscillator, defined by the Lagrangian
(note in this section, $H$ does not include the source terms)
\be
L=iy^\dagger.\dot y-H(y,y^\dagger)-Ky^\dagger-K^*y.\label{holagrange}
\ee
As in the preceding section, the action principle says
\be
\delta_K\langle 0,t_1|0,t_2\rangle^K=-i\langle 0,t_1|\int_{t_2}^{t_1}dt\,
[\delta Ky^\dagger+\delta K^*y]|0,t_2\rangle^K,
\ee
or for $t_2<t<t_1$,
\begin{subequations}
\bea
i\frac{\delta}{\delta K(t)}\langle 0,t_1|0,t_2\rangle^K=\langle 0,t_1|
y^\dagger(t)|0,t_2\rangle^K,\\
i\frac{\delta}{\delta K^*(t)}\langle 0,t_1|0,t_2\rangle^K=\langle 0,t_1|
y(t)|0,t_2\rangle^K,
\eea
\end{subequations}
where we have introduced the concept of the functional derivative.
The equation of motion
\be
i\dot y-\frac{\partial H}{\partial y^\dagger}-K=0,
\quad -i\dot y^\dagger-\frac{\partial H}{\partial y}-K^*=0,
\label{rep:eom}
\ee
is thus equivalent to the functional differential equation,
\be
0=\left\{i\left[K(t),W\left[i\frac{\delta}{\delta K^*},i\frac{\delta}{\delta K}
\right]\right]-K(t)\right\}\langle0,t_1|0,t_2\rangle^K,
\label{fnalde}
\ee
where (the square brackets indicate functional dependence)
\be
W[y,y^\dagger]=\int_{t_2}^{t_1}dt\,
[iy^\dagger(t).\dot y(t)-H(y(t),y^\dagger(t))].
\ee
The reason Eq.~(\ref{fnalde}) holds is that by definition
\be
\frac{\delta}{\delta K(t)}K(t')=\delta(t-t'),
\ee
so
\bea
&&i\left[K(t),\int_{t_2}^{t_1}dt'\left(i\frac{i\delta}{\delta K(t')}.
\frac{d}{dt'}
\frac{i\delta}{\delta K^*(t')}-H\left(\frac{i\delta}{\delta K^*(t')},
\frac{i\delta}{\delta K(t')}\right)\right)\right]\nonumber\\ 
&&\qquad=i\frac{d}{dt}\frac{i\delta}{\delta K^*(t)}-\frac\partial{\partial
(i\delta /\delta K(t))}H\left(\frac{i\delta}{\delta K^*(t)},\frac{i\delta}
{\delta K(t)}\right),
\eea
which corresponds to the first two terms in the equation of motion
(\ref{rep:eom}), under
the correspondence
\be
y\leftrightarrow i\frac\delta{\delta K^*},\quad y^\dagger\leftrightarrow
i\frac\delta{\delta K}.
\ee
Since $[[K,W],W]=0$, we can write the functional equation (\ref{fnalde}) as
\be
0=e^{iW[i\delta/\delta K^*,i\delta/\delta K]}K
e^{-iW[i\delta/\delta K^*,i\delta/\delta K]}\langle0,t_1|0,t_2\rangle^K.
\ee
The above equation has a solution (up to a constant), because both
equations (\ref{rep:eom}) must hold,
\be
\langle0,t_1|0,t_2\rangle^K=e^{iW[i\delta/\delta K^*,i\delta/\delta K]}
\delta[K]\delta[K^*],
\ee
where $\delta[K]$, $\delta[K^*]$ are functional delta functions.  
The latter have functional Fourier decompositions (up to a multiplicative
constant),
\begin{subequations}
\bea
\delta[K]&=&\int[dy^\dagger]e^{-i\int dt \,K(t)y^\dagger(t)},\\
\delta[K^*]&=&\int[dy]e^{-i\int dt \,K^*(t)y(t)},
\eea
\end{subequations}
where $[dy]$ represents an element of integration over all (numerical-valued)
{\it functions\/} $y(t)$,
and so we finally have
\bea
&&\langle0,t_1|0,t_2\rangle^{K,K^*}\nonumber\\
&&\quad=\int[dy][dy^\dagger]\exp\left(-i\int_{t_2}
^{t_1}dt\left[K(t)y^\dagger (t)+K^*(t)y(t)\right]+iW[y,y^\dagger]\right)
\nonumber\\
&&\quad=
\int[dy][dy^\dagger]\exp\left(i\int_{t_2}^{t_1}dt\left[iy^\dagger \dot y
-H(y,y^\dagger)-Ky^\dagger-K^*y\right]\right),
\label{hopathint}
\eea
where $y$, $y^\dagger$ are now numerical, and the functional integration is
over all possible functions, over all possible ``paths.''  Of course, the
classical paths, the ones for which $W-\int dt(Ky^\dagger+K^*y)$ is
an extremum, receive the greatest weight, at least in the classical limit,
where $\hbar\to0$.

\subsection{Example} 
\label{sec:hogf}
Consider the harmonic oscillator Hamiltonian,
$H=\omega y^\dagger y$.  Suppose we wish to calculate, once again, the
ground state persistence amplitude, $\langle 0,t_1|0,t_2\rangle^K$.
It is perhaps easiest to perform a Fourier transform,
\be
y(\nu)=\int_{-\infty}^\infty dt\, e^{i\nu t}y(t),\quad
 y^*(-\nu)=\int_{-\infty}^\infty dt\, e^{-i\nu t}y^\dagger(t).
\ee
Then
\begin{subequations}
\bea
\int_{-\infty}^\infty dt\,y^\dagger(t)y(t)&=&\int_{-\infty}^\infty\frac{d\nu}
{2\pi}y(\nu)y^*(-\nu),\\
\int_{-\infty}^\infty dt\,iy^\dagger(t)\dot y(t)
&=&\int_{-\infty}^\infty\frac{d\nu}{2\pi}\nu y(\nu)y^*(-\nu).
\eea
\end{subequations}
Thus Eq.~(\ref{hopathint}) becomes
\bea
&&\langle0,t_1|0,t_2\rangle^{K,K^*}\nonumber\\
&=&\int[dy][dy^*]\exp\left\{i\int
\frac{d\nu}{2\pi}\left[y(\nu)(\nu-\omega)y^*(-\nu)-y^*(-\nu)K(\nu)
-y(\nu)K^*(-\nu)\right]\right\}\nonumber\\
&=&\int[dy][dy^*]\exp\bigg\{i\int\frac{d\nu}{2\pi}\left[y(\nu)-\frac{K(\nu)}
{\nu-\omega}\right](\nu-\omega)\left[y^*(-\nu)-\frac{K^*(-\nu)}{\nu-\omega}
\right]\nonumber\\
&&\qquad\mbox{}-i\int\frac{d\nu}{2\pi}K(\nu)\frac1{\nu-\omega}K^*(-\nu)\bigg\}
\nonumber\\
&=&\int[dy][dy^*]\exp\left\{i\int\frac{d\nu}{2\pi}y(\nu)(\nu-\omega)y^*(-\nu)
\right\}\nonumber\\
&&\qquad\times
\exp\left\{-i\int\frac{d\nu}{2\pi}K(\nu)\frac1{\nu-\omega}K^*(-\nu)
\right\}
\nonumber
\\
&=&\exp\left\{-i\int\frac{d\nu}{2\pi}K(\nu)\frac1{\nu-\omega}K^*(-\nu)\right\},
\eea
since the first exponential in the penultimate line,
obtained by shifting the integration variable,
\begin{subequations}
\bea
y(\nu)-\frac{K(\nu)}{\nu-\omega}&\to& y(\nu),\\
y^*(-\nu)-\frac{K^*(-\nu)}{\nu-\omega}&\to& y^*(-\nu),
\eea
\end{subequations}
 is $\langle0,t_1|0,t_2\rangle^{K=K^*=0}=1$.
How do we interpret the singularity
at $\nu=\omega$ in the remaining integral?  We should
have inserted a convergence factor in the original functional integral:
\be
\exp\left(i\int\frac{d\nu}{2\pi}\left[\dots\right]\right)\to\exp\left(
i\int\frac{d\nu}{2\pi}\left[\dots+i\epsilon y(\nu)y^*(-\nu)\right]\right),
\ee
where $\epsilon$ goes to zero through positive values.  Thus we have, in 
effect, $\nu-\omega\to\nu-\omega+i\epsilon$ and so we have for the ground-state
persistence amplitude
\be
\langle0,t_1|0,t_2\rangle^{K,K^*}=e^{-i\int dt\,dt'\,K^*(t)G(t-t')K(t')},
\label{hovpa}
\ee
which has the form of Eq.~(\ref{gspa}), with
\be
G(t-t')=\int_{-\infty}^\infty \frac{d\nu}{2\pi}\frac{e^{-i\nu(t-t')}}{\nu
-\omega+i\epsilon},
\ee
which is evaluated by closing the $\nu$ contour in the upper half plane
if $t-t'<0$, and in the lower half plane when $t-t'>0$.  
Since the pole is in the lower half plane we get
\be
G(t-t')=-i\eta(t-t')e^{-i\omega(t-t')},
\ee
which is exactly what we found in Eq.~(\ref{hogf}).

Now, let us rewrite the path integral
(\ref{hopathint}) in terms of co\"ordinates and momenta:
\begin{subequations}
\bea
q&=&\frac1{\sqrt{2\omega}}(y+y^\dagger),\quad p=\sqrt{\frac\omega2}\frac1i
(y-y^\dagger),\\
y&=&\sqrt{\frac\omega2}\left(q+\frac{ip}\omega\right),\quad
y^\dagger=\sqrt{\frac\omega2}\left(q-\frac{ip}\omega\right).
\eea
\end{subequations}
Then the numerical Lagrangian appearing in (\ref{hopathint})
 may be rewritten as
\bea
L&=&iy^\dagger \dot y-\omega y^\dagger y-Ky^\dagger-K^*y\nonumber\\
&=&i\frac\omega2\left(q-i\frac{p}\omega\right)\left(\dot q+i\frac{\dot p}\omega
\right)-\frac{\omega^2}2\left(q^2+\frac{p^2}{\omega^2}\right)
\nonumber\\
&&\qquad\mbox{}
-\sqrt{\frac\omega2}K\left(q-\frac{ip}\omega\right)
-\sqrt{\frac\omega2}K^*\left(q+\frac{ip}\omega\right)\nonumber\\
&=&i\frac\omega4\frac{d}{dt}\left(q^2+\frac{p^2}{\omega^2}\right)+p\dot q
-\frac12\frac{d}{dt}(pq)-\frac12(p^2+\omega^2q^2)-\sqrt{2\omega}\Re Kq
-\sqrt{\frac2\omega}\Im Kp\nonumber\\
&=&\frac{d}{dt}w+L(q,\dot q,t),
\label{transholag}
\eea
where, if we set $\dot q=p$, the Lagrangian is
\be
L(q,\dot q,t)=\frac12\dot q^2-\frac12\omega^2q^2+Fq,
\label{tradlag}
\ee
if
\be
\Im K=0,\quad F=-\sqrt{2\omega}\Re K.
\ee
In the path integral
\be
[dy][dy^\dagger]
=[dq][dp]\left|\frac{\partial(y,y^\dagger)}{\partial(q,p)}\right|,
\ee
where the Jacobian is
\be
\left|\frac{\partial(y,y^\dagger)}{\partial(q,p)}\right|
=\left|\begin{array}{cc}
\sqrt{\frac\omega2}\quad&\sqrt{\frac\omega2}\\
\\
\frac{i}{\sqrt{2\omega}}\quad&-\frac{i}{\sqrt{2\omega}}\end{array}
\right|=1,
\ee
and so from the penultimate line of Eq.~(\ref{transholag}), the path integral
(\ref{hopathint}) becomes
\bea
&&\langle 0,t_1|0,t_2\rangle^F
=\int[dy][dy^\dagger]
\exp\left[i\int_{t_2}^{t_1}dt\,L(y,y^\dagger)\right]\nonumber\\
&&\quad=\int[dq][dp]\exp\left[i\int_{t_2}^{t_1}dt\left(p\dot q-\frac12p^2
-\frac12\omega^2q^2+Fq\right)\right].
\eea
Now we can carry out the $p$ integration, since it is Gaussian:
\bea
\int[dp]e^{i\int dt\left[-\frac12p^2+p\dot q\right]}
&=&\int[dp]e^{i\int dt\left[-\frac12(p-\dot q)^2+\frac12\dot q^2\right]}
\nonumber\\
&=&e^{i\int dt\frac12\dot q^2}\prod_i\int_{-\infty}^\infty dp_i \,e^{-\frac12
ip_i^2\Delta t}.
\eea
Here we have discretized time so that $p(t_i)=p_i$, so the final functional
integral over $p$ is just an infinite product of constants, each one of which
equals $e^{-i\pi/4}\sqrt{2\pi/\Delta t}$.  
Thus we arrive at the form originally
written down by Feynman \cite{Feynman1965},
\be
\langle 0,t_1|0,t_2\rangle^F=\int[dq]\exp\left\{i\int_{t_2}^{t_1}dt\,
L(q,\dot q,t)\right\},
\label{feynpathint}
\ee
with the Lagrangian given by Eq.~(\ref{tradlag}),
where an infinite normalization constant has been absorbed into the measure.